  \documentclass[journal,onecolumn,draftcls,12pt]{IEEEtran}
\usepackage{cite}
\usepackage{multirow}
\usepackage{graphicx}
\usepackage{amssymb,amsmath,times,wasysym}
\usepackage{pb-diagram}
\usepackage{psfrag,balance}
\usepackage{srcltx,verbatim,color} 
\usepackage[caption=false]{subfig} 

\usepackage[margin=0.75in]{geometry}
\usepackage{algorithmic}
\usepackage{algorithm}
\usepackage[ansinew]{inputenc} 
\usepackage{amsthm}

\newcounter{appendx}

\newcommand{\conj}[1]{\left.\!{#1}\!\right.\!^*\!}
\newcommand{\real}[1]{\Re\left({#1}\right)}

\renewcommand{\log}[2][]{\mathrm{log}_{#1}\left(#2\right)}

\newcommand{\E}[2][]{\mathbb{E}_{#1}\!\left[{#2}\right]}

\newcommand{\Var}[1]{\mathbb{V} \mathrm{ar}\!\left[{#1}\right]}

\newcommand{\Cov}[1]{\mathbb{C} \mathrm{ov}\!\left({#1}\right)}

\title{Exploiting   Underlay Spectrum Sharing  in Cell-Free   Massive 	MIMO   Systems}

\author{{Diluka Loku Galappaththige,  \IEEEmembership{Student Member, IEEE}  and  Gayan Amarasuriya Aruma Baduge, \IEEEmembership{Senior Member, IEEE}  \vspace{-10mm}}

\thanks{Authors are with the School of Electrical, Computer, and Biomedical Engineering, Southern Illinois University, Carbondale, IL, USA 62901, Email: \{diluka.lg, gayan.baduge\}@siu.edu.  This work in part has been presented  at IEEE International Conference on Communications (ICC), May 2019, Shanghai, China, \cite{Galappaththige2019} and also in IEEE ICC May  2020, Dublin, Ireland \cite{Galappaththige2020}. 
}\vspace{-8mm}}

\begin{document}
\bstctlcite{IEEEexample:BSTcontrol}
\vspace{-5mm}
\maketitle
 
\begin{abstract}
We investigate the coexistence of underlay spectrum sharing in cell-free massive multiple-input multiple-output (MIMO) systems. 
A primary system with geographically distributed primary access points (P-APs) serves a multitude of  primary users (PUs), while a secondary system serves a large number of secondary users (SUs) in the same primary/licensed spectrum by exploiting the underlay spectrum sharing.   To mitigate the secondary co-channel interference  inflected at  PUs, stringent secondary transmit power constraints are defined for the secondary access points (S-APs). A generalized pilots sharing scheme is used to locally estimate the uplink  channels at P-APs/S-APs, and thereby, conjugate precoders are adopted to serve PUs/SUs in the same time-frequency resource element. 
Moreover, the effect of a user-centric AP clustering scheme  is investigated  by assigning a suitable set of APs to a particular user. 
The impact of  estimated  downlink (DL)  channels at  PUs/SUs via DL pilots beamformed by P-APs/S-APs is  investigated. The achievable primary/secondary rates at PUs/SUs are derived for the statistical DL and estimated DL CSI cases.  User-fairness for PUs/SUs is achieved by  designing  efficient transmit power control policies based on a multi-objective optimization problem formulation of joint underlay spectrum sharing and max-min criteria. The proposed orthogonal multiple-access based analytical framework is also extended to facilitate non-orthogonal multiple-access. 
Our analysis and numerical results manifest that the primary/secondary performance  of  underlay spectrum sharing  can be boosted by virtue of the  average reduction of  transmit powers/path-losses, uniform coverage/service, and macro-diversity gains, which are inherent to distributed transmissions/receptions of cell-free massive MIMO.

\end{abstract}

\linespread{0}
\vspace{0mm}
\section{Introduction}\label{sec:introduction}
Massive multiple-input multiple-output (MIMO)   operating in sub-6\,GHz  can simultaneously serve many users in the same time-frequency resource element by virtue of aggressive spatial multiplexing gains rendered by large base-station (BS) antenna arrays \cite{Galappaththige2019,Galappaththige2020,Marzetta2010,Larsson2014}. Co-located massive MIMO in which all BS antennas are   packed into the same array is currently being deployed   in the United States \cite{Sprint}. Thus, the co-located massive MIMO enabled with fully-digital beamforming  has already become a reality \cite{Emil2019a}. 

Recently, a distributed/cell-free massive MIMO architecture, which deploys  a large number of   distributed access points (APs), is proposed to enhance the coverage probability and hence to provide a uniformly better service to users   within a much larger geographical area \cite{Ngo2015,Ngo2017,Nayebi2017,Mai2018,Doan2017,Li2018,Li2019,Zhang2020}. 
Coordinated multi-point (CoMP) and network MIMO \cite{Hosseini2014} are two related technologies, which exploit the notion of  cooperation among BSs located within dedicated  cells. On the contrary to   these existing techniques, all APs are deployed in  a cell-free architecture  aiming to jointly serve all users  for a given/large geographical area with no cell-boundaries. 
The uplink (UL) channel state information (CSI) is locally estimated at each AP via pilots send by users, and these APs are connected to a central processing unit (CPU) via  a fronthaul/backhaul network \cite{Ngo2017}.  
By exploiting the channel reciprocity of time-division duplexing (TDD) mode of operation, APs acquire the downlink (DL) CSI via UL channel estimates and thereby design precoders for DL data transmission.  The requirement for CSI exchange among  APs via the CPU strictly depends on the AP precoder design. 
Owing to the fact that the distributed APs enable a user-centric architecture, the average transmission distances of a cell-free massive MIMO are inherently  smaller than that of the co-located counterpart \cite{Dai2011}. This benefit translates into transmit power and path-loss reductions, which in turn  lead to boosted energy efficiencies.
The distributed APs circumvent the impediments caused by  spatially correlated fading and shadowing due to large obstacles. 
The underlying macro-diversity gains can be exploited to boost the achievable rates \cite{Koyuncu2018,Liu2020}.
Joint beamfoming at a massive number of APs also enables unprecedented spatial multiplexing gains. Thus, cell-free massive MIMO enables a large  number of concurrent connections  with a guaranteed uniform service throughout a given/large geographical area. 

Cognitive radio techniques based on the spectrum sharing are extensively explored to mitigate  spectrum scarcity and underutilization/holes for the next-generation wireless communication systems \cite{Goldsmith2009,Baduge2018,Wang2017}. 
In particular, many spectrum sharing techniques are evolved through three main  paradigms, namely underlay, overlay, and interweave \cite{Goldsmith2009}. The overlay   spectrum sharing  involves sophisticated signal processing techniques and requires  codebook knowledge of the non-cognitive users. The interweave spectrum sharing adopts opportunistic frequency reuse over the available spectrum holes and hence requires stringent activity information of the non-cognitive users. The  underlay spectrum sharing allows cognitive/secondary users (SUs) to simultaneously operate within the licensed spectrum if the secondary interference inflicted at the non-cognitive/primary users (PUs) is below a certain threshold. The main reasons for adopting the underlay spectrum sharing  in this paper are its implementation  simplicity  and the achievable high spectrum utilization \cite{Goldsmith2009,Baduge2018,Wang2017}, compared to the significantly sophisticated overlay  and interweave counterparts.	
Thus, in the underlay spectrum sharing, an underlaid secondary system can be simultaneously operated in the same licensed primary spectrum  by defining stringent secondary transmit power constraints such that the secondary co-channel interference (CCI) caused to the  PUs always falls below a predefined primary interference threshold (PIT) \cite{Goldsmith2009}.

\subsection{Related prior research on underlay spectrum sharing with   massive MIMO}\label{sec:literature}	
In \cite{Wang2017}, an initial foundation for investigating the feasibility of underlay spectrum sharing  in massive MIMO with co-located antenna arrays at BSs is established. The BSs in \cite{Wang2017} use the maximum ratio transmission (MRT) for signal transmission, and the DL achievable rate of the secondary system which is underlaid in a primary massive MIMO system is investigated. 
Reference \cite{Baduge2018} investigates the impact  of inherent intra/inter-cell pilot contamination in multi-cell multi-user massive MIMO system with underlay spectrum sharing. In \cite{Li2017}, a dual-hop enabled spectrum sharing system is analyzed, and the achievable rates are derived.  Thereby, the detrimental effects of inter/intra-cell pilot contamination are investigated.  
The fundamental performance limits for relay selection strategies in massive MIMO two-way relaying are explored for  perfect CSI in \cite{Silva2020}. Moreover, in \cite{Silva2020}, the asymptotic achievable rates are derived for the best relay selection by deriving the asymptotic signal-to-interference-plus-noise ratio (SINR).
In \cite{Kudathanthirige2019}, the achievable rates of reserve-TDD based underlay spectrum sharing   are presented. In \cite{Li2019}, the achievable rates of underlay spectrum sharing co-located massive MIMO non-orthogonal multiple-access (NOMA) are presented. 
In \cite{Rezaei2020},   a low-complexity sub-optimal user-clustering  technique for  NOMA based underlay spectrum sharing in cell-free massive MIMO  is proposed, and thereby, the achievable rates are derived for fixed transmit power allocation. 
In \cite{Kusaladharma2017,Kusaladharma2016}, the performance bounds of spectrum sharing for massive MIMO   with stochastic BS/user locations are derived. Moreover, by investigating pilot contamination, path-loss inversion power control, and receiver association policies,  the secondary interference for a random cognitive massive MIMO system is characterized in \cite{Kusaladharma2017,Kusaladharma2016}. 
In \cite{Chaudhari2018}, the quality-of-service aware power allocation and user selection schemes are studied for cognitive massive MIMO systems. 
Pilot  decontamination techniques are proposed to asymptotically mitigate the residual interference in an underlaid single user massive MIMO cognitive radio system in \cite{Filippou2012}. 
In \cite{Koyuncu2018},  the macro-multiplexing gain achieved from  optimization of antenna locations is characterized in terms of the ambient dimension  of the cell and the path-loss exponent.
Reference \cite{Liu2020} derives the upper and lower bounds of the achievable  rate with the perfect/imperfect CSI for cell-free massive MIMO systems. Thereby, \cite{Liu2020} shows that the bounds of the achievable rate converge to a common lower bound owing to the  extra distance-diversity or macro-diversity gain offered by distributed antennas in cell-free massive MIMO. 	
In \cite{Masoumi2020}, the effects of finite capacity of fronthauls   in the presence of  residual hardware impairments at the users and APs are investigated by deriving the achievable   rates for the compress-forward-estimate, estimate-compress-forward, and  estimate-multiply-compress-forward strategies. Reference \cite{Alonzo2019} proposes  a low-complexity power allocation scheme to maximize the energy efficiency for a cell-free massive MIMO system with user centric approach operating at millimeter-wave (mmWave) frequencies. In \cite{Chen2020},  two pilot assignments, namely the user-group  and interference-based $K$-means schemes are proposed for the structured massive access.

\subsection{Our motivation}\label{sec:motivation}	
The aforementioned related prior references \cite{Baduge2018,Li2017,Kudathanthirige2019,Silva2020,Kusaladharma2017,Filippou2012,Wang2017} have investigated the coexistence of massive MIMO and underlay spectrum sharing with only co-located antenna arrays at the BSs. 
The closely related  references \cite{Li2019} and \cite{Rezaei2020}, respectively, 
investigate the rate performance of  co-located and cell-free massive MIMO NOMA with underlay spectrum sharing by only considering fixed power allocation.
To the best of our knowledge, multi-objective max-min fairness-based transmit power allocation, impact of beamformed DL pilots by APs, and the achievable rates with DL estimated CSI at   users have not yet been investigated. Thus, in our paper, we fill this gap by exploring  multi-objective power allocation, impact of DL  pilots,  and  practically realizable performance bounds of underlay spectrum sharing in cell-free massive MIMO with imperfectly estimated UL/DL CSI. 
Since the APs are spatially distributed over a given geographical area, the distributed transmissions of cell-free massive MIMO architecture can be beneficial in providing   uniformly better average rate to users by virtue of mitigating the near-far effect via max-min power control than that of the co-located massive MIMO counterparts. The average amount of secondary CCI inflicted at a particular PU from cell-free/distributed massive MIMO  transmissions  can be better constrained as specified by the PIT with low-power distributed APs.
Moreover, cell-free massive MIMO has more robustness against the detrimental effects of correlated small/large scale fading than the co-located counterpart \cite{Ngo2017}. Both primary and secondary systems can provide a higher coverage   probability because there are no cell boundaries, and users are much closer to the APs in cell-free setting.
Thus, the cell-free massive MIMO can significantly boost  the performance of underlay spectrum sharing.

\subsection{Our contribution and its difference relative to the existing literature}\label{sec:contribution}
Our main contribution is to investigate  different UL/DL CSI cases at APs/users, and   transmit power control and their effects on the achievable rates of  underlay spectrum sharing in orthogonal multiple-access (OMA)/NOMA-based cell-free massive MIMO. 
Specifically, we derive the   performance metrics  by proposing max-min based multi-objective transmit power control algorithms and by exploring the  deleterious impact of imperfectly estimated UL CSI at APs,  availability of estimated DL CSI at   the users, and effects of using statistical DL CSI with imperfect successive interference cancellation (SIC) signal decoding.     
Both the primary system and the secondary system, which is underlaid within the primary licensed spectrum,  adopt a generalized pilot sharing scheme  to minimize the training overhead.
Thus, the pilot sequences sent by PUs and SUs are used to locally estimate the UL channels at primary access points (P-APs)/secondary access points (S-APs), respectively. Thereby, the impact of imperfectly estimated CSI is considered for our analysis. Our performance metrics for cell-free underlay spectrum sharing are categorized by taking into account the availability of long-term/statistical CSI and estimated DL CSI at  PUs/SUs. When only UL channel estimation is used, the users rely on statistical CSI for signal decoding. Nonetheless, when APs beamform DL pilots, the users adopt estimated DL CSI to decode signals. Thus, the performance bounds are established for these two user CSI cases. Moreover, the effect of user-centric achieved through clustering the APs that serve for a particular user in both primary an secondary systems is investigated for cell-free   underlay spectrum sharing. 

To mitigate the secondary CCI inflected at  PUs due to simultaneous transmission, the stringent secondary transmit power constraints are introduced at  S-APs.
Thus, by defining a PIT at  PUs, the secondary transmit power is constrained such that the total secondary CCI at any PU falls bellow the predefined PIT.  Then, in the presence of imperfectly estimated UL CSI at APs and DL CSI at users with intra-system pilot contamination and estimation errors, the achievable rates for  PUs/SUs are derived. 
A multi-objective transmit power control algorithm is designed based on the max-min fairness criterion to guarantee a uniform quality-of-service among all the users.
Moreover, the above OMA-based system model is extended to facilitate  NOMA transmissions, and the corresponding performance bounds are established. 
The practical viability of underlay spectrum sharing in cell-free massive MIMO is explored by using numerical results through our analysis and Monte-Carlo simulations.  

This paper goes well beyond our related conference papers  \cite{Galappaththige2019,Galappaththige2020} by presenting a multi-objective optimization of   max-min fairness-based transmit power allocation for the OMA-based primary system,  user-centric AP clustering  aspects,  beamforming of DL pilots,  impact of estimated DL channel estimates, adoption of estimated DL CSI for signal decoding,  and the corresponding achievable rates with estimated DL CSI  for underlay spectrum sharing within    cell-free massive MIMO.  All numerical results/comparisons, except for Fig. \ref{fig:rate_vs_No_US_NOMA_OMA}, and their descriptions  are distinctive from those of \cite{Galappaththige2019,Galappaththige2020}, and the corresponding figures have been regenerated for different system parameters with respect to \cite{Galappaththige2019,Galappaththige2020}. 
 
\noindent \textbf{Notation:}      $\mathbf z^{T}$  denotes the transpose of $\mathbf z$ . The conjugate of $z$ is denoted by $z^*$. The notation $z\sim \mathcal {CN}(\cdot,\cdot)$ denotes that $z$ is a complex-valued circularly symmetric Gaussian distributed random variable. The operators $\E{\cdot}$ and $\Var{\cdot}$ are the expectation and variance, respectively.

\section{System, channel and signal models  }\label{sec:system_model}
\subsection{System and channel model for OMA/NOMA}\label{sec:system_and_channel}

\begin{figure}[!t]\centering \vspace{-4mm}
	\def\svgwidth{440pt} 
		\fontsize{9}{7}\selectfont 
	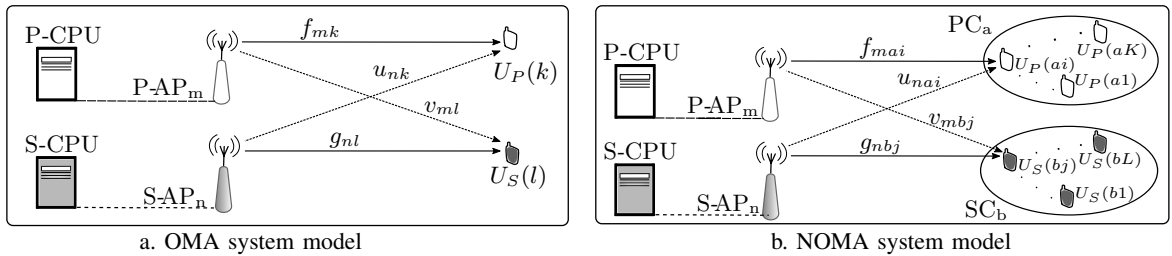 \vspace{-2mm}
	\caption{A   cell-free   massive MIMO system models with underlay spectrum sharing.}\vspace{-10mm} \label{fig:system_model}
\end{figure}

We consider a TDD cell-free massive MIMO system with underlay spectrum sharing (see Fig. \ref{fig:system_model}a). 
A secondary system is underlaid within a primary licensed spectrum in  order to enhance the overall spectrum efficiency by eliminating spectrum holes.
The primary system having $M$ single-antenna P-APs serves $K$ single-antenna PUs, while the secondary system with $N$ single-antenna S-APs uses the same time-frequency spectrum to serve $L$ single-antenna SUs. 
We introduce  the secondary transmit power constraints for all S-APs to ensure that the performance of the primary system is not hindered by the simultaneous secondary transmission in the same primary   spectrum.
Thus, the  CCI caused by the secondary system at  PUs falls below a predefined PIT, which defines a upper  limit for the CCI endurance ability of PUs.  The PIT constraint mitigates excessive  secondary CCI at PUs.
A synchronized operation between the primary and the secondary systems is assumed such that all P-APs and S-APs simultaneously serve all PUs and SUs by adopting spatial multiplexing rendered by cell-free massive MIMO \cite{Ngo2017}.
Moreover, P-APs/S-APs are connected to their respective  primary/secondary central processing units (P-CPU/S-CPU). The $k$th PU and the $l$th SU are denoted by $U_P(k)$ and $U_S(l)$, respectively.

In Fig. \ref{fig:system_model}a,  $f_{mk}$, $g_{nl}$, $v_{ml}$, and $u_{nk}$ are the channel coefficients between  the $m$th P-AP and $U_P(k)$, the $n$th S-AP and $U_S(l)$, the $m$th P-AP and $U_S(l)$, and the $n$th S-AP and $U_P(k)$, respectively, where $m \in \{1, \cdots, M\}$, $k\in\{1, \cdots, K\}$, $n\in\{1, \cdots, N\}$, and $l\in\{1, \cdots, L\}$.
The above channels can be modeled in a unified manner as
\vspace{-1mm}
\begin{eqnarray}\label{eqn:Channel_rep}
	{ h}_{ab}  =  \tilde { h}_{ab}  {{\zeta}^{1/2}_{{ h}_{ab}}},
\end{eqnarray} 

\vspace{-1mm}
\noindent where $h\in\{f,g,u,v\}$, $a\in\{m,n\}$, and $b\in\{k,l\}$. Here, ${\tilde { h}_{ab} \sim \mathcal {CN}(0,1)}$ captures the independent quasi-static Rayleigh fading and stays fixed during the coherence interval, while  $\zeta_{{ h}_{ab}}$ accounts for the large-scale fading, including path-loss and shadow fading. Since  the large-scale coefficients stay fixed for several coherence intervals,  it is assumed that the large-scale coefficients are known a-prior at both P-APs and S-APs \cite{Ngo2017}. Thus, the estimation of these large-scale fading coefficients can be done in once about tens/hundreds of coherence intervals \cite{Marzetta2016_Book}. 

Next, we extend our   cell-free massive MIMO underlay spectrum sharing techniques to  facilitate NOMA. To this end, we consider a system setup with $AK$ number of PUs and $BL$ number of SUs, where $A$ and $B$ are the numbers of primary clusters (PCs) and secondary clusters (SCs), respectively (see Fig. \ref{fig:system_model}b). 
Based on the spatial directions of the users \cite{Kudathanthirige2019,Ding2016},  $K$ and $L$ number of PUs and SUs are assigned to each PC and SC, respectively.
The $k$th PU in the $a$th PC and the $l$th SU in the $b$th SC are denoted by $U_P(a,k)$ and $U_S(b,l)$, respectively. 
The channel  between the $m$th P-AP and $U_P(a,k)$ is denoted by $f_{mak}$, where $m \in \{1,\cdots,M \}$, $a \in \{1,\cdots,A\}$, and $k \in \{1,\cdots,K\}$. The channel between the $n$th S-AP and $U_S(b,l)$ is represented by $g_{nbl}$, where $n \in \{1,\cdots,N \}$, $b \in \{1,\cdots,B\}$, and $l \in \{1,\cdots,L\}$. Further, $v_{mbl}$ and $u_{nak}$ denote the interference channels between the $m$th P-AP and $U_S(b,l)$, and the $n$th S-AP and $U_P(a,k)$, respectively. These channels are modeled   similar to  \eqref{eqn:Channel_rep} as 
\vspace{-1mm}
\begin{eqnarray}\label{eqn:Channel_rep_noma}
	{ h}_{qrs}  =  \tilde { h}_{qrs}  {{\zeta}^{1/2}_{{ h}_{qrs}}},
\end{eqnarray}

\vspace{-1mm}
\noindent where $h \in \{f,g,u,v\}$, $q \in \{m,n\}$, $r \in \{a,b\}$, and $s \in \{k,l\}$. Moreover, $\tilde { h}_{qrs} \sim \mathcal{CN}(0,1)$ captures the small-scale fading, while   ${{\zeta}_{{ h}_{qrs}}}$ captures the large-scale fading, including path-loss and shadowing.

\subsection{The UL channel estimation for OMA/NOMA}\label{sec:Channel state information acquisition}
For OMA, the UL channels are estimated locally at P-APs and S-APs by using respective user pilots \cite{Ngo2017}. During UL channel estimation period, $\tau_p$ symbols out of the coherence interval having $\tau_c$ symbols are used to transmit the UL pilot. Then, by using these pilot sequences, the channels $f_{mk}$ and $g_{nl}$ are estimated at  P-APs and S-APs, respectively.  
In practice, the number of orthogonal pilot sequences  is limited as it is defined  by the channel coherence interval \cite{Marzetta2016_Book}. To reduce the pilot overhead and to increase the number of served PUs/SUs,  in this work, the pilot sequences are shared among  PUs and SUs by  adopting the  following  pilot sharing strategy.
It is assumed that $Q \!\leq\! \min\!\left(\!K,L\!\right)$ number of pilot sequences having a length of $\tau_p$ symbol duration is shared among PUs and SUs. Then, we define the complete pilot sets used by  PUs $(\mathbf \Phi_{P})$  and SUs  $(\mathbf \Phi_{S})$  as 
\vspace{-1mm}
\begin{eqnarray}\label{eqn:pilot_sequences}
	\mathbf \Phi_{P} = \left[\mathbf \Phi ; \tilde{ \mathbf \Phi}_{P}\right]  \qquad \text{and} \qquad \mathbf \Phi_{S} = \left[\mathbf \Phi ; \tilde{ \mathbf \Phi}_{S}\right],
\end{eqnarray}

\vspace{-1mm}
\noindent where $\mathbf \Phi \in \mathbb C^{Q\times\tau_{p}}$ denotes the pilots   shared by $Q$ PUs/SUs  having a length of $\tau_{p}$ symbol duration.  $\tilde{ \mathbf \Phi}_{P} \in \mathbb C^{\left(K-Q\right)\times\tau_{p}}$, and $\tilde{ \mathbf \Phi}_{S} \in \mathbb C^{\left(L-Q\right)\times\tau_{p}}$ are the pilots assigned for the remaining $\left(K-Q\right)$ PUs or $\left(L-Q\right)$ SUs, respectively. We can define the orthogonal properties among these pilots   as $\mathbf \Phi^H \tilde{ \mathbf \Phi}_{P}= 0$, $\mathbf \Phi^H\tilde{ \mathbf \Phi}_{S}= 0$, and $\tilde{ \mathbf \Phi}_{P}^H\tilde{ \mathbf \Phi}_{S}= 0$. We define $\mathbf \Phi_{P} = \left[\boldsymbol{\phi}_{P_1}^T,\cdots, \boldsymbol{\phi}_{P_k}^T, \cdots, \boldsymbol{\phi}_{P_K}^T \right]^T$ and  $\mathbf \Phi_{S} = \left[\boldsymbol {\phi}_{S_1}^T,\cdots, \boldsymbol {\phi}_{S_l}^T,  \cdots, \boldsymbol {\phi}_{S_L}^T   \right]^T$,  where  $\boldsymbol{\phi}_{P_k} \in \mathbb C^{1\times\tau_{p}}$ 
and $\boldsymbol {\phi}_{S_l} \in \mathbb C^{1\times\tau_{p}}$ are the  pilot sequences sent by   $U_P(k)$  and $U_S(l)$, respectively, and $\|\boldsymbol {\phi}_{P_k}\|^2=1$ and $\|\boldsymbol {\phi}_{S_l}\|^2=1$ for $k\in\{1, \cdots, K\}$ and $l\in\{1, \cdots, L\}$.
Then, we can write the pilot signal received at the $m$th P-AP  and the $n$th S-AP as 
\vspace{-1mm}
\begin{subequations}
	\begin{eqnarray}
	 	\mathbf y'_{P_m}  &=& \sqrt{P_p}  \sum\nolimits_{k=1}^{K} f_{mk} \boldsymbol {\phi}_{P_k} + \sqrt{P_p}  \sum\nolimits_{l=1}^{L} v_{ml}\boldsymbol {\phi}_{S_l} +  \mathbf n_{P_m}',
		\label{eqn:rx_pilot_signal_at_m_PAP}\\
	 	\mathbf y'_{S_n}  &= &\sqrt{P_p}  \sum\nolimits_{l=1}^{L} g_{nl} \boldsymbol {\phi}_{S_l} + \sqrt{P_p}  \sum\nolimits_{k=1}^{K} u_{nk} \boldsymbol {\phi}_{P_k} +  \mathbf n_{S_n}',
		\label{eqn:rx_pilot_signal_at_n_SAP}
	\end{eqnarray}
\end{subequations} 

\vspace{-1mm}
\noindent where $P_p = \tau_{p} P$ and $P$ denotes the average transmitted pilot power at each PU/SU. Moreover, $\mathbf n_{P_m}'$ and $\mathbf n_{S_n}'$ are additive white Gaussian noise (AWGN) vectors, having independent and identically distributed  (i.i.d.) $\mathcal {CN}(0,1)$ elements  at the $m$th P-AP and the $n$th S-AP, respectively.
The sufficient statistics for estimating $f_{mk}$ and $g_{nl}$ can be obtained by projecting $\boldsymbol {\phi}_{P_k}^H$ and $\boldsymbol {\phi}_{S_l}^H$ onto \eqref{eqn:rx_pilot_signal_at_m_PAP} and \eqref{eqn:rx_pilot_signal_at_n_SAP}, respectively, as 	
\vspace{-1mm}
\begin{subequations}
	\begin{eqnarray}
		 y_{P_{mk}}  &=& \boldsymbol \phi_{P_k}^H \mathbf y'_{P_m}  = \sqrt{P_p} f_{mk} + \sqrt{P_p} v_{mk} +    n_{P_m}, \label{eqn:est_signl_m_PAP}\\
		 y_{S_{nl}}  &=&  \boldsymbol \phi_{S_l}^H \mathbf y'_{S_n}  = \sqrt{P_p} g_{nl} + \sqrt{P_p} u_{nl} +    n_{S_n}, \label{eqn:est_signl_n_SAP}
	\end{eqnarray}
\end{subequations}

\vspace{-1mm}
\noindent where $\{\phi_{P_k},\phi_{S_l}\} \!\in \!{\bf \Phi}$, $ n_{P_m} \!\!=\! \boldsymbol \phi_{P_k}^H \mathbf n_{P_m}' \!\sim\! \mathcal {CN}(0,1) $,  and $ n_{S_n} \!\!=\! \boldsymbol \phi_{S_l}^H \mathbf n_{S_n}' \!\sim\! \mathcal {CN}(0,1) $  as $\boldsymbol \phi_{P_k}$ and $\boldsymbol \phi_{S_l}$ are unitary vectors. 

\noindent \textbf{\textit{Proposition 1:}}
\textit{The minimum mean square error (MMSE) estimates of $f_{mk}$ and $g_{nl}$    
are given by} \vspace{-1mm}
\begin{subequations}\label{eqn:estimate_channel}
	\begin{eqnarray}
		\hat {f}_{mk} &=& \left({\E {y_{P_{mk}}^* f_{mk}}}\big/{\E {|y_{P_{mk}}|^2}}\right)  y_{P_{mk}} = c_{P_{mk}} y_{P_{mk}} ,\label{eqn:estimate_f_mk}\\
		\hat {g}_{nl} &=& \left({\E {y_{S_{nl}}^* g_{nl}}}\big/{\E {|y_{S_{nl}}|^2}}\right)  y_{S_{nl}} = c_{S_{nl}} y_{S_{nl}}\label{eqn:estimate_g_nl},
	\end{eqnarray}
\end{subequations}

\vspace{-1mm}
\noindent \textit{where $c_{P_{mk}}$ and $c_{S_{nl}}$ are given by}
\vspace{-1mm}
\begin{eqnarray}
	c_{P_{mk}} &=& \frac{\sqrt{\tau_{p} P} {\zeta}_{{f}_{mk}}}{\tau_{p} P \left({\zeta}_{{f}_{mk}} + {\zeta}_{{v}_{mk}}\right)+1}  \label{eqn:cpmk} \qquad \text{and}\qquad
	c_{S_{nl}} = \frac{\sqrt{\tau_{p} P} {\zeta}_{{g}_{nl}}}{\tau_{p} P \left({\zeta}_{{g}_{nl}} + {\zeta}_{{u}_{nl}}\right)+1}.\label{eqn:csnl}
\end{eqnarray}
\vspace{-6mm}
\begin{proof}
	  Appendix \ref{app:Appendix1}.
\end{proof}

\noindent  \textbf{\textit{Remark} 1}: The MMSE channel estimates in \eqref{eqn:estimate_f_mk} and \eqref{eqn:estimate_g_nl}  are valid for the coexistence of PUs/SUs with shared pilot sequences defined by (\ref{eqn:pilot_sequences}).  The  MMSE channel estimation with the conventional orthogonal pilots for  cell-free massive MIMO is reported in \cite{Ngo2017}.

Due to  TDD channel reciprocity,   P-APs and S-APs utilize locally estimated $\hat {f}_{mk}$ and $\hat {g}_{nl}$ as    DL CSI to construct their precoders  \cite{Marzetta2016_Book}. Furthermore, the actual channels can be written as
\vspace{-1mm}
\begin{eqnarray}
	{f}_{mk} &=& \hat {f}_{mk} + \epsilon_{f_{mk}}  \label{eqn:actual_f_mk} \qquad \text{and}\qquad 
	{g}_{nl}   =  \hat {g}_{nl} + \epsilon_{g_{nl}},\label{eqn:actual_g_nl}
\end{eqnarray}

\vspace{-1mm}
\noindent where $\epsilon_{f_{mk}}$ and $\epsilon_{g_{nl}}$ are the estimation errors, which are independent of the corresponding channel estimates yielded from orthogonality property of MMSE criterion \cite{Kay1993}.

For  NOMA, the UL channel $(f_{mak})$ is also estimated locally at  P-APs  from the pilots   sent by PUs within PCs.
Again, we assume that $Q \!\leq\! \min (A,B)$ number of pilots   is shared among  PCs and SCs as  
\vspace{-1mm}
\begin{eqnarray}\label{eqn:noma_pilot_sequences}
	\mathbf \Phi_{PA} = \left[\mathbf \Phi ; \tilde{ \mathbf \Phi}_{PA}\right] \qquad \text{and} \qquad \mathbf \Phi_{SB} = \left[\mathbf \Phi ; \tilde{ \mathbf \Phi}_{SB}\right],
\end{eqnarray}

\vspace{-1mm}
\noindent where $\mathbf \Phi \in \mathbb C^{Q\times\tau_{p}}$ denotes the shared $Q$-pilot sequence among PCs and SCs. Then, the remaining $(A-Q)$ PCs or $(B-Q)$ SCs are assigned with the pilot sequences $\tilde{ \mathbf \Phi}_{PA} \in \mathbb C^{(Q-A)\times\tau_{p}}$ and $\tilde{ \mathbf \Phi}_{SB} \in \mathbb C^{(Q-B)\times\tau_{p}}$, respectively. Furthermore,  $\mathbf \Phi_{PA} = \left[\boldsymbol{\phi}_{P_1}^T,\cdots, \boldsymbol{\phi}_{P_a}^T, \cdots, \boldsymbol{\phi}_{P_A}^T \right]^T$ and  $\mathbf \Phi_{SB} = \left[\boldsymbol {\phi}_{S_1}^T,\cdots, \boldsymbol {\phi}_{S_b}^T,  \cdots, \boldsymbol {\phi}_{S_B}^T   \right]^T$, where $\boldsymbol{\phi}_{P_a} \in \mathbb C^{1\times\tau_{p}}$ and $\boldsymbol{\phi}_{S_b} \in \mathbb C^{1\times\tau_{p}}$ are the pilot sequences assigned to the $a$th PC and the $b$th SC, satisfying  $\| \boldsymbol{\phi}_{P_a} \|^2 = 1$ and $\| \boldsymbol{\phi}_{S_b} \|^2 = 1$ for $a \in \{1,\cdots,A\}$ and $b \in \{1, \cdots,B\}$.
The received pilot vector at the $m$th P-AP can be written as
\vspace{-1mm}
\begin{eqnarray}
	\mathbf y_{P_m}  &=& \sqrt{P_p} \sum\nolimits_{a=1}^{A} \sum\nolimits_{k=1}^{K} f_{mak} \boldsymbol {\phi}_{P_a} + \sqrt{P_p}  \sum\nolimits_{b=1}^{B}\sum\nolimits_{l=1}^{L} v_{mbl}\boldsymbol {\phi}_{S_b} +  \mathbf n_{P_m}, \label{eqn:noma_rx_pilot_m_PAP}
\end{eqnarray}

\vspace{-1mm}
\noindent where $\mathbf n_{P_m} \sim \mathcal {CN}(0,1)$ is the AWGN vector at the $m$th P-AP. Then, by projecting $\boldsymbol {\phi}_{P_a}^H \in {\bf \Phi}$  into \eqref{eqn:noma_rx_pilot_m_PAP}, we obtain a sufficient statistic  to estimate $f_{mak}$  as
\vspace{-1mm}
\begin{eqnarray}
	y_{P_{ma}}  &=& \boldsymbol \phi_{P_a}^H \mathbf y_{P_m}  = \sqrt{P_p} \sum\nolimits_{i=1}^{K}f_{mak} + \sqrt{P_p} \sum\nolimits_{j=1}^{L} v_{maj} +    n_{P_m}. \label{eqn:noma_est_m_PAP}
\end{eqnarray}

\vspace{-1mm}
\noindent By following steps similar to those in Appendix \ref{app:Appendix1}, the MMSE estimate of $f_{mak}$ can be derived as 
\vspace{-1mm}
\begin{eqnarray}
	\hat {f}_{mak} &=& \frac{\E {y_{P_{ma}}^* f_{mak}}}{\E {|y_{P_{ma}}|^2}}  y_{P_{ma}} = \frac{\sqrt{\tau_{p} P} {\zeta}_{{f}_{mak}}}{\tau_{p} P \left( \sum_{i=1}^{K}{\zeta}_{{f}_{mai}} + \sum_{j=1}^{L}{\zeta}_{{v}_{maj}}\right)+1} y_{P_{ma}}.\label{eqn:noma_est_f_mak}
\end{eqnarray}

\vspace{-1mm}
\noindent Since $y_{P_{ma}}$ is Gaussian distributed, we have $\hat {f}_{mak} \sim \mathcal{CN}(0,\alpha_{f_{mak}})$, where $\alpha_{f_{mak}}$   is given as
\vspace{-1mm}
\begin{eqnarray}
	\alpha_{f_{mak}} &=& \E{|\hat f_{mak}|^2} = \frac{\tau_{p} P \zeta_{{f}_{mak}}^2}{ \tau_{p} P \left(\sum_{i=1}^{K} \zeta_{{f}_{mai}} + \sum_{j=1}^{L} \zeta_{{v}_{maj}} \right) + 1}. \label{eqn:theta_f}
\end{eqnarray}

\vspace{-1mm}
\noindent Then, the channel estimation error of $f_{mak}$ is defined as $\epsilon_{f_{mak}} = f_{mak} - \hat f_{mak} \sim \mathcal{CN}(0,\zeta_{{f}_{mak}}-\alpha_{f_{mak}})$.

\noindent  \textbf{\textit{Remark} 2}:
The parameters $	y_{S_{nb}} $,  $\hat {g}_{nbl}$, and $\alpha_{g_{nbl}}$  corresponding to the secondary  system can be obtained by replacing subscripts $\{P,m, a, K, k, i\}$  of \eqref{eqn:noma_est_m_PAP}, \eqref{eqn:noma_est_f_mak},  and \eqref{eqn:theta_f}, respectively, by  $\{S, n, b, L, l, j\}$. Hence, the explicit presentation of UL channel estimates at S-APs is omitted for the sake of brevity.

\subsection{AP Clustering for OMA}\label{sec:AP_clustering}
The P-APs/S-APs are clustered to serve a particular set of users based on the locally estimated channel gains. We assume that only $\mathcal{M}_{P} = \{1, \cdots, M_P\}$ and  $\mathcal{N}_{S} = \{1, \cdots, N_S\}$ sets of P-APs/S-APs are assigned to  $U_P(k)$ and $U_S(l)$, respectively \cite{Emil2019}.
At each P-AP/S-AP, a set of transmission determination coefficients  can be introduced based on the  estimated channels   such that $\delta_{P_{mk}}=\delta_{S_{nl}}=1$ for $m \in \mathcal{M}_P$, $n \in \mathcal{N}_{S}$, and $\delta_{P_{mk}}=\delta_{S_{nl}}=0$, otherwise.  

\noindent  \textbf{\textit{Remark} 3}: When the aforementioned AP clustering is adopted for the proposed cell-free underlay spectrum sharing,  the primary/secondary CCI can be  reduced, and hence, the achievable rates of both primary/secondary systems can be boosted with respect to a conventional unclustered system as depicted in Fig. \ref{fig:sum_rate_USC_32_16} in our numerical results in Section \ref{sec:Numerical}.

\subsection{Signal model for OMA}\label{sec:Signal model}
Due to  implementation simplicity and near optimal performance in large AP regime \cite{Marzetta2016_Book}, conjugate precoding is used at    P-APs and S-APs to transmit  signals towards their respective users via the channel estimates in \eqref{eqn:estimate_f_mk} and \eqref{eqn:estimate_g_nl}.
We  can write the transmitted signals at the $m$th P-AP and the $n$th S-AP as
\vspace{-1mm}
\begin{eqnarray}
	{x}_{P_m}  &=&  \sqrt{P_P}   \sum\nolimits_{k=1}^{K} \delta_{P_{mk}}   {\eta_{P_{mk}}^{1/2} \hat {f}_{mk}^* q_{P_k}} \label{eqn:mth transmit signal}  \quad \text{and}\quad 
	{x}_{S_n}   =   \sqrt{P_{S}} \sum\nolimits_{l=1}^{L}  \delta_{S_{nl}} {\eta_{S_{nl}}^{1/2} \hat {g}_{nl}^* q_{S_l}}, \label{eqn:nth transmit signal}
\end{eqnarray}

\vspace{-1mm}
\noindent where $\eta_{P_{mk}}$ and $\eta_{S_{nl}}$ are the  power allocation coefficients at the $m$th P-AP and the $n$th S-AP, respectively. Moreover, $P_P$ and $P_S$ denote the maximum allowable transmit powers at each P-AP and S-AP, respectively.
Here, $\eta_{S_{nl}}$ is selected to satisfy the total transmit power constraints given in \eqref{eqn:SAP_tx_pwr}. The signals intended to  $U_P(k)$ and  $U_S(l)$ are denoted by $q_{P_k}$ and $q_{S_l}$, respectively, and they satisfy $\E {|q_{P_k}|^2} = 1$ and $\E {|q_{S_l}|^2} = 1$. 
Then, the received signals at  $U_P(k)$ and $U_S(l)$ can be written as
\vspace{-1mm}
\begin{eqnarray}
 	\!\!\!\!\!\!\!\!\!\! {r}_{P_k} &=& \sum\nolimits_{m=1}^{M} {f_{mk} {x}_{P_m}} + 	\sum\nolimits_{n=1}^{N} {u_{nk} {x}_{S_n}} + n_{P_k} \label{eqn:k_PU_Rx_signal}
	\quad \text{and}\quad
	{r}_{S_l} =\sum\nolimits_{n=1}^{N} {g_{nl} {x}_{S_n}} + \sum\nolimits_{m=1}^{M} {v_{ml} {x}_{P_m}} + n_{S_l},\label{eqn:l_SU_Rx_signal}
\end{eqnarray}

\vspace{-1mm}
\noindent where $n_{P_k}\sim \mathcal{CN}(0,1)$ and $n_{S_l}\sim \mathcal{CN}(0,1)$ are    AWGN at  $U_P(k)$ and  $U_S(l)$, respectively.  Here, ${x}_{P_m}$ and $ {x}_{S_n}$ are given in (\ref{eqn:mth transmit signal}).
We can rearrange the received signal at  $U_P(k)$ in \eqref{eqn:k_PU_Rx_signal}  as
\vspace{-1mm} 
\begin{eqnarray}\label{eqn:k_PU_Rx_signal_reagd}
	{r}_{P_k} &=&  \sqrt{P_{P}}  \sum\nolimits_{m=1}^{M}  \delta_{P_{mk}}  {\eta_{P_{mk}}^{1/2} f_{mk} \hat{f}_{mk}^* q_{P_k}}  
	+   \sqrt{P_{P}}  \sum\nolimits_{m=1}^{M}   \sum\nolimits_{i\neq k}^{K}  \delta_{P_{mi}}  {\eta_{P_{mi}}^{1/2} f_{mk} \hat{f}_{mi}^* q_{P_i}}  \nonumber \\
	&&+  \sqrt{P_S} \sum\nolimits_{n=1}^{N} \sum\nolimits_{j=1}^{L}  \delta_{S_{nj}} {\eta_{S_{nj}}^{1/2} u_{nk} \hat{g}_{nj}^* q_{S_j}} + n_{P_k},
\end{eqnarray} 

\vspace{-1mm}
\noindent where the first term represents the desired signal component at  $U_P(k)$, while the inter-system interference caused by beamforming uncertainty of conjugate precoding is captured by the second term. The  third term accounts for the intra-system interference yielded from the secondary CCI.
Similarly, the received signal at $U_S(l)$ in \eqref{eqn:l_SU_Rx_signal} can be rewritten as
\vspace{-1mm}
\begin{eqnarray}\label{eqn:l_SU_Rx_signal_reagd}
	{r}_{S_l} &=&  \sqrt{P_{S}}  \sum\nolimits_{n=1}^{N} \delta_{S_{nl}} {\eta_{S_{nl}}^{1/2} g_{nl} \hat{g}_{nl}^* q_{S_l}}   +   \sqrt{P_{S}}  \sum\nolimits_{n=1}^{N}  \sum\nolimits_{j\neq l}^{L} \delta_{S_{nj}} {\eta_{S_{nj}}^{1/2} g_{nl} \hat{g}_{nj}^* q_{S_j}}   \nonumber \\
	&&+  \sqrt{P_P} \sum\nolimits_{m=1}^{M}  \sum\nolimits_{i=1}^{K} \delta_{P_{mi}} {\eta_{P_{mi}}^{1/2} v_{ml} \hat{f}_{mi}^* q_{P_i}} + n_{S_l}.
\end{eqnarray}

\subsection{Signal model for NOMA}\label{sec:noma_Signal model}
Again, P-APs  employ conjugate precoding  by using the channel estimates in \eqref{eqn:noma_est_f_mak}. The transmitted signal at the $m$th P-AP  can be written  as
\vspace{-1mm}
\begin{eqnarray}
	{x}_{P_m}  &=&  \sqrt{P_P} \sum\nolimits_{a=1}^{A}  \sum\nolimits_{i=1}^{K}   {\eta_{P_{ma i}}^{1/2} \hat {f}_{m a i}^* q_{P_{a i}}},   \label{eqn:noma_mth transmit signal} 
\end{eqnarray}

\vspace{-1mm}
\noindent where $\eta_{P_{ma i}}$ is the transmit power control coefficient at the $m$th P-AP. Here,   $\E{|q_{P_{a i}}|^2}=1$, where $q_{P_{a i}}$ is the signal intended for $U_P(a,i)$. Thus,   the received signal at $U_P(a,k)$ can be written  as
\vspace{-1mm}
\begin{eqnarray}\label{eqn:noma_k_PU_Rx_signal}
 	{r}_{P_{ak}} &=&  \sqrt{P_{P}}  \sum\nolimits_{m=1}^{M}   {\eta_{P_{mak}}^{1/2} f_{mak} \hat{f}_{mak}^* q_{P_{ak}}}  
	+   \sqrt{P_{P}}  \sum\nolimits_{m=1}^{M}   \sum\nolimits_{i\neq k}^{K}  {\eta_{P_{mai}}^{1/2} f_{mak} \hat{f}_{mai}^* q_{P_{ai}}}   \\
	&&\!\!\!\!\!\!\!\!\!\!\!\!\!\! + \! \sqrt{P_P} \sum\nolimits_{m=1}^{M} \! \sum\nolimits_{a' \neq a}^{A}\! \sum\nolimits_{i=1}^{K} \!\eta_{P_{ma' i}}^{1/2} f_{mak} \hat{f}_{ma' i}^* q_{P_{a' i}}  \!+\! \sqrt{P_S} \sum\nolimits_{n=1}^{N}\! \sum\nolimits_{b =1}^{B} \! \sum\nolimits_{j=1}^{L} \! {\eta_{S_{nb j}}^{1/2} u_{nak} \hat{g}_{nb j}^* q_{S_{b j}}} \!+\! n_{P_{ak}}, \nonumber
\end{eqnarray} 

\vspace{-1mm}
\noindent where $n_{P_{ak}} \sim \mathcal{CN}(0,1)$ is the AWGN at $U_P(a,k)$. 
To apply the power-domain NOMA, we assume that the users in the $a$th PC are  ordered based on the effective channel strength as   \cite{Kudathanthirige2019,Li2018,Ding2016}
\vspace{-1mm}
\begin{eqnarray}\label{eqn:noma_eff_chn_strngth}
	\E{\left| \sum\nolimits_{m=1}^{M} \hat{f}_{ma1} \right|^2} \geq \cdots \geq \E{\left| \sum\nolimits_{m=1}^{M} \hat{f}_{mak} \right|^2} \geq \cdots \geq \E{\left| \sum\nolimits_{m=1}^{M} \hat{f}_{maK} \right|^2}.
\end{eqnarray} 

\vspace{-1mm}
\noindent In power-domain NOMA,  higher transmit powers are allocated for the users with weaker channel conditions. Thus, the transmit powers are ordered as  \cite{Kudathanthirige2019,Li2018,Ding2016}
\vspace{-1mm}
\begin{eqnarray}\label{eqn:noma_powrs}
	P_{P_{a1}} \leq \cdots \leq P_{P_{ak}} \leq \cdots \leq P_{P_{aK}},
\end{eqnarray}

\vspace{-1mm}
\noindent where $P_{P_{ak}} = P_P \eta_{P_{mak}}$.
Consequently, $U_P(a,k)$ aims to decode the signal intended for $U_P(a,i)$ for $\forall i \geq k$ provided that  $U_P(a,k)$ can decode its own signal. Hence, $U_P(a,k)$ may  successively cancel the intra-cluster interference from $U_P(a,i)$ before decoding   its own signal for $\forall i  \geq k$, and the residual interference due to SIC error propagation must also captured. Moreover, $U_P(a,k)$ treats the signals for   users $\forall i < k$ as interference  \cite{Kudathanthirige2019,Li2018,Ding2016}.
To this end, the received signal at $U_P(a,k)$ upon imperfect SIC with error propagation can be written as 
\vspace{-1mm}
\begin{eqnarray}\label{eqn:noma_PRx_signal_aftr_sic}
	{r}_{P_{ak}} &=& \underbrace{\sqrt{P_{P}}  \sum\nolimits_{m=1}^{M} \eta_{P_{mak}}^{1/2} f_{mak} \hat{f}_{mak}^* q_{P_{ak}} }_{\text{Desired signal}} + \underbrace{ \sqrt{P_P} \sum\nolimits_{m=1}^{M} \sum\nolimits_{i=1}^{k-1} \eta_{P_{mai}}^{1/2} f_{mai} \hat{f}_{mai}^* q_{P_{ai}} }_{\text{Intra-cluster interference after SIC}} \\
	&&\!\!\!\!\!\!\!\!\!\!\!\!+ \underbrace{ \sqrt{P_P} \sum\nolimits_{m=1}^{M} \sum\nolimits_{i=k+1}^{K} \eta_{P_{mai}} \left(f_{mak} \hat{f}_{mai}^* q_{P_{ai}} - \E{f_{mak} \hat{f}_{mai}^* }\hat{q}_{P_{ai}} \right) }_{\text{Error propagation due to imperfect SIC}} \nonumber \\  
	&&\!\!\!\!\!\!\!\!\!\!\!\!+ \! \underbrace{ \sqrt{P_P} \!\sum\nolimits_{m=1}^{M}\! \sum\nolimits_{a' \neq a}^{A} \!\sum\nolimits_{i =1}^{K}\! \eta_{P_{m a' i}}^{1/2} f_{mak} \hat{f}_{m a' i}^* q_{P_{a' i}} }_{\text{Intra-system interference}} \!+\! \underbrace{ \sqrt{P_S}\! \sum\nolimits_{n=1}^{N}\! \sum\nolimits_{b =1}^{B} \!\sum\nolimits_{j =1}^{L}\! \eta_{S_{n b j}}^{1/2} u_{nak} \hat{g}_{n b j}^* q_{S_{b j}} }_{\text{Inter-system interference}} \!+\! \underbrace{n_{P_{ak}}}_{\text{AWGN}}.\nonumber
\end{eqnarray}

\noindent  \textbf{\textit{Remark} 4}: In our proposed cell-free  NOMA-aided underlay spectrum sharing, the perfect SIC   is not feasible due to intra-cluster pilot contamination, intra-system interference, inter-system interference, channel estimation errors, and statistical CSI knowledge at the users. Thus,    the residual interference caused by imperfect SIC needs to be modeled. The third term in (\ref{eqn:noma_PRx_signal_aftr_sic}) captures the error propagation due to imperfect SIC in which  $\hat{q}_{P_{ai}}$ is the estimate of ${q}_{P_{ai}}$. Since ${q}_{P_{ai}}$ is Gaussian distributed, $\hat{q}_{P_{ai}}$ and ${q}_{P_{ai}}$ are assumed to be jointly Gaussian distributed with a normalized correlation coefficient $\vartheta_{P_{ai}}$ as \cite{Li2018}
\vspace{-1mm}
\begin{eqnarray}\label{eqn:noma_est_signl}
	{q}_{P_{ai}} = \vartheta_{P_{ai}} \hat{q}_{P_{ai}} + e_{P_{ai}},
\end{eqnarray}

\vspace{-1mm}
\noindent where $\hat{q}_{P_{ai}} \sim \mathcal{CN}(0,1)$, $ e_{P_{ai}} \sim \mathcal{CN}(0,\sigma_{e_{P_{ai}}}^2 / (1+ \sigma_{e_{P_{ai}}}^2))$, and $\vartheta_{P_{ai}} = 1/\sqrt{1+ \sigma_{e_{P_{ai}}}^2}$. Furthermore, $\hat{q}_{P_{ai}}$ and $e_{P_{ai}}$ are statistically independent. Thus, the third term in (\ref{eqn:noma_PRx_signal_aftr_sic})  can be used to capture the residual interference caused by error propagation of imperfect SIC when evaluating the  SINR and  achievable rate.

\noindent  \textbf{\textit{Remark} 5}:
The received signal at $U_S(b,l)$ for the secondary  system  $({r}_{S_{bl}})$ and the signal intended for $(U_S(b,j)$, ${q}_{S_{bj}})$ can be obtained by replacing the subscripts $\{P, M, m, A, a, K, k, i\}$ of \eqref{eqn:noma_PRx_signal_aftr_sic} and \eqref{eqn:noma_est_signl}, respectively,  by   $\{S, N, n, B, b, L, l, j\}$.

\subsection{Secondary Transmit Power Constraints for OMA}\label{sec:Secondary Transmit Power Constraints}
We constrain the transmit power at each S-AP to guarantee that the secondary CCI inflected at  PUs falls below the   PIT of each  PU. Thus, we define the total transmit power at the $n$th S-AP as  
\vspace{-1mm}
\begin{eqnarray}
	P_{S_n} \triangleq \sum\nolimits_{l=1}^{L} P_{S} \delta_{S_{nl}} \eta_{S_{nl}} \quad \text{for}  \quad n\in\{1, \cdots, N\} \quad \text{and} \quad l\in\{1, \cdots, L\},  \label{S_total_tx_power}
\end{eqnarray}

\vspace{-1mm}
\noindent where  $\sum_{l=1}^{L} \delta_{S_{nl}} \eta_{S_{nl}}\leq 1 $. Moreover, the transmit power allocation coefficient at the  $n$th S-AP for $U_S(l)$ is represented by $\eta_{S_{nl}}$.
Thus, the CCI  received at  $U_P(k)$ from  all S-APs can be written as
\vspace{-1mm}
\begin{eqnarray}\label{eqn:Rx_k PU_from_SAPs}
	y_{k} &=& \sum\nolimits_{n=1}^{N}  u_{nk} x_{S_{n}}  = \sqrt{P_{S}}\sum\nolimits_{n=1}^{N} \sum\nolimits_{l=1}^{L} \delta_{S_{nl}} \eta_{S_{nl}}^{1/2} u_{nk} \hat {g}_{nl}^* q_{S_l},
\end{eqnarray}

\vspace{-1mm}
\noindent where $x_{S_n}$ is the transmitted signal at the $n$th S-AP and   defined in (\ref{eqn:nth transmit signal}). 

\noindent \textbf{\textit{Proposition 2:}}
\textit{The total average secondary CCI power ($P_{I_k}$) inflicted    at  $U_P(k)$ is given by  }
\vspace{-1mm}
\begin{eqnarray}\label{eqn:rx_interf_power}
	 P_{I_k} &=& \E {\left| y_{k} \right|^2} = {P_{S}}\underbrace{\sum\nolimits_{n=1}^{N} \sum\nolimits_{l=1}^{L} \delta_{S_{nl}}  \eta_{S_{nl}}   \E {\left|  {u_{nk} \hat {g}_{nl}^* } \right|^2}}_{Z_k},
\end{eqnarray}

\vspace{-1mm}
\noindent \textit{where  $Z_k$  is defined as}
\vspace{-1mm}
 \begin{eqnarray}\label{eqn:Z_k}
	Z_k &=& \sum\nolimits_{n=1}^{N} \sum\nolimits_{l=1}^{L} \delta_{S_{nl}} \eta_{S_{nl}} \rho_{g_{nl}} \zeta_{{u}_{nk}} + \sum\nolimits_{n=1}^{N} \delta_{S_{nk}} \eta_{S_{nk}} \rho_{u_{nk}}^2,
\end{eqnarray}

\vspace{-1mm}
\noindent \textit{where  $\rho_{g_{nl}} \triangleq \sqrt{\tau_{p} P} c_{S_{nl}} \zeta_{{g}_{nl}} $ and $\rho_{u_{nk}} \triangleq \sqrt{\tau_{p} P} c_{S_{nk}} \zeta_{{u}_{nk}}$. }
\begin{proof}
	  Appendix \ref{app:Appendix2}.
\end{proof}

\vspace{-1mm}
\noindent  Thus, we can give the secondary transmit power constraint at the $n$th S-AP as follows:
\vspace{-1mm}
\begin{eqnarray}\label{eqn:SAP_tx_pwr}
	P_{S_n} = \min{\left(P_{S}, {I_{T_1}}\big/{Z_{1}}, \cdots, {I_{T_k}}\big/{Z_{k}}, \cdots, {I_{T_K}}\big/{Z_{K}} \right)},
\end{eqnarray}

\vspace{-1mm}
\noindent where $I_{T_k}$ is the interference threshold of  $U_P(k)$.

\subsection{Secondary transmit power control for NOMA}\label{sec:noma_S_tx_power_cntrol}
As per  Section \ref{sec:Secondary Transmit Power Constraints}, to ensure that the performance of  primary system is not hindered by the secondary system, we constrain the transmit power of S-APs. The secondary CCI received at $U_P(a,k)$ is given by
\vspace{-8mm}
\begin{eqnarray}\label{eqn:noma_rx_kPU_from_SAPs}
	y_{ak} &=& \sum\nolimits_{n=1}^{N}  u_{nak} x_{S_{nb}}  = \sqrt{P_{S}} \sum\nolimits_{n=1}^{N} \sum\nolimits_{b=1}^{B} \sum\nolimits_{l=1}^{L}  \eta_{S_{nbl}}^{1/2} u_{nak} \hat {g}_{nbl}^* q_{S_{bl}},
\end{eqnarray}

\vspace{-1mm}
\noindent where $x_{S_{nb}}$ is the transmit signal intended for $L$ users in the $b$th SC. Then, the total average secondary CCI   ($P_{I_{ak}}$) inflicted    at $U_P(a,k)$ can be derived  as 
\vspace{-1mm}
\begin{eqnarray}\label{eqn:noma_rx_interf_power}
	P_{I_{ak}} &=& \E {\left| y_{ak} \right|^2} = {P_{S}}\underbrace{\sum\nolimits_{n=1}^{N} \sum\nolimits_{b=1}^{B} \sum\nolimits_{l=1}^{L}   \eta_{S_{nl}}   \E {\left|  {u_{nak} \hat {g}_{nbl}^* } \right|^2}}_{Z_{ak}}, 
\end{eqnarray}

\vspace{-1mm}
\noindent where  $Z_{ak}$ can be derived by following steps  similar to those  in Appendix \ref{app:Appendix2} as
\vspace{-1mm}
\begin{eqnarray}\label{eqn:noma_Z_k}
	Z_{ak} &=& \sum\nolimits_{n=1}^{N} \sum\nolimits_{b=1}^{B} \sum\nolimits_{l=1}^{L} \eta_{S_{nbl}} \alpha_{g_{nbl}} \zeta_{{u}_{nak}} + \sum\nolimits_{n=1}^{N} \sum\nolimits_{l=1}^{L}  \eta_{S_{nal}} \alpha_{f_{mak}}^2 \left(\frac{\zeta_{{u}_{uak}} \zeta_{{g}_{nal}}}{\zeta_{{f}_{mak}}} \right)^2.
\end{eqnarray}

\vspace{-1mm}
\noindent Thus, we derive the secondary transmit power constraint at the $n$th S-AP  as
\vspace{-1mm}
\begin{eqnarray}\label{eqn:noma_SAP_tx_pwr}
	P_{S_n} = \min{\left(P_{S}, {I_{T_{11}}}\big/{Z_{11}}, \cdots, {I_{T_{ak}}}\big/{Z_{ak}}, \cdots, {I_{T_{AK}}}\big/{Z_{AK}} \right)},
\end{eqnarray}

\vspace{-1mm}
\noindent where $I_{T_{ak}}$ is the interference threshold of $U_P(a,k)$.

\section{Achievable Rate analysis of OMA-Aided Underlay Spectrum Sharing}\label{sec:Achievable Sum rates Definitions}
\subsection{Achievable rate analysis for the primary system}\label{sec:Achievable sum rate analysis for primary system}
When the  P-APs/S-APs do not beamform DL pilots for the acquisition of DL CSI,  the PUs/SUs are unaware of  the instantaneous DL channel coefficients.
Thus, the PUs/SUs must relay on the long-term/statistical DL  channel coefficients for signal detection \cite{Marzetta2016_Book}. This is a typical scenario in TDD-based co-located massive MIMO in which only UL pilots are used to estimate UL channels at the P-APs/S-APs, and the instantaneous DL channel coefficients can be approximated by their statistical counterparts thanks to channel hardening property.     
To this end, the signal received at  $U_P(k)$ can be rearranged as 
\vspace{-1mm}
\begin{eqnarray}\label{eqn:k_PU_WG_Rx_signal}
	\!\!\!\! r_{P_k} \!&=&\! \underbrace{\sqrt {P_{P}} \E {\sum\nolimits_{m=1}^{M}\!  \delta_{P_{mk}} {\eta_{P_{mk}}^{1/2} f_{mk} \hat{f}_{mk}^*}} q_{P_k} }_{\text{Desired signal}}  \nonumber\\ 
	&&\!\!\!\!+ \underbrace{\sqrt{P_{P}}  \left( {\sum\nolimits_{m=1}^{M} \! \delta_{P_{mk}} {\eta_{P_{mk}}^{1/2} f_{mk} \hat{f}_{mk}^* } } \!-\!  {\E{\sum\nolimits_{m=1}^{M} \! \delta_{P_{mk}} {\eta_{P_{mk}}^{1/2} f_{mk} \hat{f}_{mk}^* } }}\right) q_{P_k} }_{\text{Detection uncertainty}}  \nonumber\\ 
	&&\!\!\!\!+\!\underbrace{\sqrt {P_{P}} \sum\nolimits_{i \neq k}^{K}{\sum\nolimits_{m=1}^{M} \delta_{P_{mi}} {\eta_{P_{mi}}^{1/2} f_{mk} \hat{f}_{mi}^* } }  q_{P_i}}_{\text{Inter-system interference caused by beamforminmg uncertainty}} \!+\! \underbrace{\sqrt{P_{S}}\sum\nolimits_{j=1}^{L} {\sum\nolimits_{n=1}^{N} \delta_{S_{nj}} {\eta_{S_{nj}}^{1/2} u_{nk} \hat{g}_{nj}^* } }  q_{S_j}}_{\text{Inter-system interference caused by secondary CCI}} \!+\! \underbrace{n_{P_k}}_{\text{AWGN}}\!.
\end{eqnarray}

\vspace{-1mm}
\noindent In (\ref{eqn:k_PU_WG_Rx_signal}),  the effective noise  consists of the sum of (i) interference caused by  detection uncertainty, (ii) intra-system interference due to beamforming uncertainty, (iii) inter-system interference due to secondary CCI, and (iv) AWGN. The desired signal term and effective noise     are uncorrelated. 
Due to the law of large number, the latter can be treated as worst-case independently distributed Gaussian noise \cite{Ngo2017n,Marzetta2016_Book}. 
\noindent \textbf{\textit{Theorem 1}}: \textit{The effective SINR at  $U_P(k)$ is given by}
\vspace{-1mm}
\begin{eqnarray}\label{eqn:SINR_ k_PU}
	\gamma_{P_k}  =  \frac{{P_{P}} \left| \E {\sum\nolimits_{m=1}^{M} \delta_{P_{mik}} {\eta_{P_{mk}}^{1/2} f_{mk} \hat{f}_{mk}^*}} \right|^2 }{P_{P}\Var{\sum\nolimits_{m=1}^{M} \delta_{P_{mk}} {\eta_{P_{mk}}^{1/2} f_{mk} \hat{f}_{mk}^*}}+ \sum\nolimits_{j=1}^{2} {I_{P}}_j + \E{\left|n_{P_k}\right|^2 }} ,
\end{eqnarray}

\vspace{-1mm}
\noindent \textit{where ${I_P}_j$ for $j \in \{1,2\}$ can be defined as}
\vspace{-1mm}
\begin{eqnarray}
	\!\!\!\!\! {I_P}_1 \!=\! {P_{P}} \E {\left|\sum\nolimits_{i \neq k}^{K} \!{\sum\nolimits_{m=1}^{M} \! \delta_{P_{mi}} {\eta_{P_{mi}}^{1/2} f_{mk} \hat{f}_{mi}^* } }\right|^2 }\!   \quad \text{and} \quad
	{I_P}_2 \!=\! {P_{S}} \E {\left|\sum\nolimits_{j=1}^{L}\! {\sum\nolimits_{n=1}^{N}\! \delta_{S_{nj}} {\eta_{S_{nj}}^{1/2} u_{nk} \hat{g}_{nj}^* } }\right|^2 }\!.{\label{eqn:Ip2} } 
\end{eqnarray} 

\vspace{-1mm}
\noindent \textit{By evaluating the expectation and variance terms in  \eqref{eqn:SINR_ k_PU} and (\ref{eqn:Ip2}),  the SINR  is given by}
\vspace{-2mm}
\begin{eqnarray}\label{eqn:sinr_k_PU}
	\!\!\!\!\!\!\!\!\!  \gamma_{P_k} \!=\! \frac{P_{P} \left(\sum\nolimits_{m=1}^{M} \delta_{P_{mk}} \eta_{P_{mk}}^{1/2} \rho_{f_{mk}} \right)^{2}} { P_{P} \!\sum\nolimits_{m=1}^{M}  \sum\nolimits_{i=1}^{K} \delta_{P_{mi}} \eta_{P_{mi}} \rho_{f_{mi}} \zeta_{f_{mk}} \!+\! {P_{S}} \!\sum\nolimits_{n=1}^{N} \sum\nolimits_{j=1}^{L} \delta_{S_{nj}} \eta_{S_{nj}} \rho_{g_{nj}} \zeta_{{u}_{nk}}  \!+\! {P_{S}} \!\sum\nolimits_{n=1}^{N} \delta_{S_{nk}} \eta_{S_{nk}} \rho_{{u}_{nk}}^2  \!+\!1 }\!,
\end{eqnarray}

\vspace{-1mm}
\noindent \text{where} $\rho_{f_{mk}} \triangleq \sqrt{\tau_{p} P} c_{P_{mk}} \zeta_{{f}_{mk}} $.
\begin{proof}
	  Appendix \ref{app:Appendix3}.
\end{proof}

\noindent
Then, we define the achievable rate of  $U_P(k)$ as follows:
\vspace{-1mm}
\begin{eqnarray}\label{eqn:rate_of_k_PU}
	R_{P_k} = \left({\left(\tau_c -\tau_p\right)}/{\tau_c}\right)\log[2] {1+ \gamma_{P_k}} ,
\end{eqnarray}

\vspace{-1mm}
\noindent where the effective portion of coherence interval for payload data transmission is captured by the pre-log factor $({\left( \tau_c -\tau_p\right)}/{\tau_c})$ and $\gamma_{P_k}$ is defined in (\ref{eqn:sinr_k_PU}).

\subsection{Achievable rate definition for the secondary system}\label{sec:Achievable sum rate analysis for secondaryy system}
By following \textbf{\textit{Theorem} 1},  we derive the achievable rate of $U_S(l)$ as follows: 
\vspace{-1mm}
\begin{eqnarray}\label{eqn:rate_ l_ SU}
	\!\!\!\!\!\!  R_{S_l} \!=\! \left({\left(\tau_c \!-\!\tau_p\right)}/{\tau_c}\right)\log[2] {1+ \gamma_{S_l} } \,\,\;\text{and}\,\,\; 	\gamma_{S_l} \!=\! \frac{{P_{S}} \left| \E {\sum\nolimits_{n=1}^{N}  \delta_{S_{nl}} {\eta_{S_{nl}}^{1/2} g_{nl} \hat{g}_{nl}^*}} \right|^2 }{P_{S}\Var{\sum\nolimits_{n=1}^{N} \delta_{S_{nl}} {\eta_{S_{nl}}^{1/2} g_{nl} \hat{g}_{nl}^*}}\!+\! \sum\nolimits_{j=1}^{2} {I_S}_j \!+\!\E{\left|w_{S_l}\right|^2} } ,\label{eqn:SINR_ l_SU}
\end{eqnarray}

\vspace{-1mm}
\noindent where  ${I_S}_j$ for $j\in\{1,2\}$ can be defined as
\vspace{-1mm}
\begin{eqnarray}
	\!\!\!\!\!\! I_{S_1} \!=\! {P_{S}} \E {\left|\sum\nolimits_{j \neq l}^{L}\!{\sum\nolimits_{n=1}^{N}\! \delta_{S_{nj}} {\eta_{S_{nj}}^{1/2} g_{nl} \hat{g}_{nj}^* } }\right|^2 } \quad \text{and} \quad
	I_{S_2} \!=\!  P_{P} \E {\left|\sum\nolimits_{i=1}^{K}\! {\sum\nolimits_{m=1}^{M}  \delta_{P_{mi}}\! \eta_{P_{mi}}^{1/2} v_{ml} \hat{f}_{mi}^* } \right|^2}.   	
\end{eqnarray}	 

\vspace{-1mm}
\noindent Thus, by deriving the expectation and variance term in \eqref{eqn:SINR_ l_SU} in similar manner to \eqref{eqn:SINR_ k_PU}, the SINR of $U_S(l)$ can be computed as 
\vspace{-1mm}
\begin{eqnarray}\label{eqn:sinr_l_SU}
	\!\!\!\!\!\!\!\!  \gamma_{S_l} \!=\! \frac{{P_{S}} \left(\sum\nolimits_{n=1}^{N}\delta_{S_{nl}}  \eta_{S_{nl}}^{1/2} \rho_{g_{nl}} \right)^{2}} { P_{S} \! \sum\nolimits_{n=1}^{N} \sum\nolimits_{j=1}^{L} \delta_{S_{nj}} \eta_{S_{nj}} \rho_{g_{nj}} \zeta_{g_{nl}} \!+\! P_{P}\! \sum\nolimits_{m=1}^{M} \sum\nolimits_{i=1}^{K} \delta_{P_{mi}} \eta_{P_{mi}} \rho_{f_{mi}} \zeta_{{v}_{ml}} \!+\! P_{P}\! \sum\nolimits_{m=1}^{M} \delta_{P_{ml}} \eta_{P_{ml}} \rho_{{v}_{ml}}^2  \!+\! 1 }\!.
\end{eqnarray}

\section{Transmit Power Control }\label{sec:Power Control}
In cell-free massive MIMO, the  user-fairness must be guaranteed in terms of the achievable rate in order to provide a uniform quality-of-service to all users. To this end, max-min power control algorithms have been shown to be optimal in the sense of user-fairness in presence of near-far effects   \cite{Marzetta2016_Book,Marbach2003,Radunovic2007,Zheng2018}. For our proposed system model, a multi-objective optimization problem (MOOP) \cite{Bjornson2014} is most suited as    both primary and secondary systems operate simultaneously. Furthermore, the secondary transmit power constraints in  \eqref{eqn:SAP_tx_pwr} must also be considered when formulating this MOOP. 

We compute the optimal power allocation coefficients of P-APs/S-APs to maximize the minimum achievable DL rate among all PUs/SUs by invoking the max-min optimization criterion \cite{Marzetta2016_Book}. 
Since the   rates in \eqref{eqn:rate_of_k_PU}  and \eqref{eqn:rate_ l_ SU} are monotonically increasing functions of their arguments, we can equivalently replace   $R_{P_k}$ and $R_{S_l}$  by $\gamma_{P_k}$ and $\gamma_{S_l}$, which are given in \eqref{eqn:sinr_k_PU} and \eqref{eqn:sinr_l_SU}, respectively. 
Thus, we formulate a max-min transmit power control problem  by   introducing a common SINR $\left(\lambda_{\gamma}\right)$ for primary/secondary systems and   by defining slack variables $\beta_{P_{mk}} \triangleq \eta_{P_{mk}}^{1/2}$ and $\beta_{S_{nl}} \triangleq \eta_{S_{nl}}^{1/2}$ as 
\vspace{-1mm}
\begin{subequations}\label{eqn:sinr_power_alloction}
	\begin{eqnarray}
	\underset{ \beta_{P_{mk}},  \beta_{S_{nl}} \forall{m,k,n,l}}{\text{maximize}} \quad&& \left(\gamma_{P_k}\right)^{w_P} \left(\gamma_{S_l}\right)^{w_S} = \lambda_{\gamma}, \label{eqn:sinre_max_min}\\
	\text{subject to}  \qquad
	&& C_1:\gamma_{P_k} \geq \lambda_{\gamma} \label{eqn:P_sinr_constraint} \quad \text{and}\quad
	C_2: \gamma_{S_l} \geq \lambda_{\gamma}, \label{eqn:S_sinr_constraint}\\
	&& C_3: \sum\nolimits_{k=1}^{K} \delta_{P_{mk}} \beta_{P_{mk}}^2 \rho_{f_{mk}} \leq 1 \quad \text{and} \quad \sum\nolimits_{l=1}^{L} \delta_{S_{nl}}  \beta_{S_{nl}}^2 \rho_{g_{nl}} \leq 1,\label{eqn:Tx_ power_constraint_ sinr_P}\\
	&& C_4:  P_{S_n} \leq {I_{T_k}}/{Z_k} ,\label{eqn:Tx_ power_constraint_ sinr_S}\\
	&& C_5: 0 \leq \beta_{P_{mk}} \quad \text{and}\quad 0 \leq \beta_{S_{nl}}, \label{eqn:power_constraint_sinr}
	\end{eqnarray}	 
\end{subequations} 

\vspace{-1mm}
\noindent where $w_P$ and $w_S$ are the priorities assigned for primary and secondary achievable SINR, respectively. Moreover, $C_3$ is obtained by invoking the maximum allowable  transmit power constraints at the $m$th P-AP and the $l$th S-AP as follows:  
\vspace{-1mm}
\begin{eqnarray}\label{eqn:derivation secondary power constrain}
	&& \E{|x_{P_m}|^2} \leq P_{P} \;\Rightarrow   \;
	\E{\left|\sqrt{P_{P}} \sum\nolimits_{k=1}^{K} \delta_{P_{mk}} {\eta_{P_{mk}}^{1/2} \hat {f}_{mk}^* q_{P_k}}\right|^2}  \leq  P_{P} \nonumber \\
	&&	\sum\nolimits_{k=1}^{K} \delta_{P_{mk}} \eta_{P_{mk}} \E{ |  \hat {f}_{mk}^* q_{P_k} |^2}  \leq  1 \;\;\Rightarrow \;\;
	\sum\nolimits_{k=1}^{K} \delta_{P_{mk}} \eta_{P_{mk}} \rho_{f_{mk}}  \leq  1.
\end{eqnarray} 

\vspace{-1mm}
\noindent It can be shown that $\sum_{l=1}^{L} \delta_{S_{nl}}  \eta_{S_{nl}} \rho_{g_{nl}}  \leq  1$. We adopt the secondary transmit power constraint in \eqref{eqn:SAP_tx_pwr} to obtain $C_4$. 
Since the objective functions in \eqref{eqn:sinr_power_alloction} are quasi-concave functions, we can show that the underlaying optimization problem is also quasi-concave \cite{Ngo2017}. Thus, an optimal solution can be found by using the Bisection method as shown in Algorithm 1. 

\begin{algorithm}\label{algorithm1}
	\caption{Bisection Algorithm}
	\begin{algorithmic}[1]
		\renewcommand{\algorithmicrequire}{\textbf{Input:}}
		\renewcommand{\algorithmicensure}{\textbf{Output:}}
		\REQUIRE Path-losses between all  P-APs/S-APs and  PUs/SUs, and the average transmit powers of  PUs/SUs.\\
		\ENSURE  The power control coefficients $\eta_{P_{mk}}$ for $m \in \{1, \cdots, M\} $, $k \in \{1, \cdots, K\}$, and $\eta_{S_{nl}}$ for $n \in \{1, \cdots, N\} $,  $l \in \{1, \cdots, L\}$. \\
		
		\textbf{Initialization}: 
		Define an initial region for the objective functions by choosing appropriate values for $\lambda_{min}$ and $\lambda_{max}$. Choose a tolerance $\epsilon > 0$.
		
		\WHILE{$ \lambda_{max} - \lambda_{min}  > \epsilon$}
		
		\STATE Calculate, $\lambda_{\gamma} = {\left(\lambda_{max}+ \lambda_{min}\right)}/{2}$. 
		
		\STATE Solve the convex feasibility problem, which can be formulated as
		\vspace{-1mm}
			\begin{eqnarray} 
			\!\!\!\!\!\!\! \| \boldsymbol{V}_{\gamma_{P_k}} \| \leq \frac{1}{\sqrt{\lambda}} \left(\sum\nolimits_{m=1}^{M} \delta_{P_{mk}} \beta_{P_{mk}} \rho_{f_{mk}}  \right) \qquad \text{and} \qquad \| \boldsymbol{V}_{\gamma_{S_l}} \| \leq \frac{1}{\sqrt{\lambda}} \left(\sum\nolimits_{n=1}^{N} \delta_{S_{nl}} \beta_{S_{nl}} \rho_{g_{nl}}  \right),
			\end{eqnarray}
		
		\vspace{-1mm}
		\noindent which is subjected to $C_2$, $C_3$, and $C_4$ given in \eqref{eqn:Tx_ power_constraint_ sinr_P}, \eqref{eqn:Tx_ power_constraint_ sinr_S}, and \eqref{eqn:power_constraint_sinr}, respectively. Moreover, $\boldsymbol{V}_{\gamma_{P_k}} \triangleq  \left[\mathbf{v}_{P_1}^T,  \; \frac{P_S}{P_P}  \mathbf{v}_{P_2}^T,  \; \frac{P_S}{P_P} \mathbf{v}_{P_3}^T,  \; \frac{1}{\sqrt{P_P}} \right]$, where 
		\vspace{-1mm}
		\begin{subequations}
		\begin{eqnarray}\label{eqn:v1}
				\!\!\!\!\!\!\!\!\!\!\!\!\!\!\!\!\!\mathbf{v}_{P_1} &\triangleq& \left[ \delta_{P_{11}} \beta_{P_{11}} \sqrt{\rho_{f_{11}} \zeta_{{f}_{1k}}} , \cdots , \delta_{P_{MK}} \beta_{P_{MK}} \sqrt{\rho_{f_{MK}} \zeta_{{f}_{Mk}}}  \, \right]^T\!\!\!, \\
				\!\!\!\!\!\!\!\!\!\!\!\!\!\!\!\!\!\mathbf{v}_{P_2} &\triangleq& \left[ \delta_{S_{11}} \beta_{S_{11}} \sqrt{\rho_{g_{11}} \zeta_{{u}_{1k}}} , \cdots , \delta_{S_{NL}} \beta_{S_{NL}} \sqrt{\rho_{g_{NL}} \zeta_{{u}_{Nk}}}  \, \right]^T\!\!\!, \\
				\!\!\!\!\!\!\!\!\!\!\!\!\!\!\!\!\!\mathbf{v}_{P_3} &\triangleq& \left[ \delta_{S_{1k}} \beta_{S_{1k}} c_{P_{1k}} \zeta_{{u}_{1k}} , \cdots , \delta_{S_{Nk}} \beta_{S_{Nk}} c_{P_{Nk}} \zeta_{{u}_{Nk}}  \, \right]^T\!\!\!,
		\end{eqnarray} 
		\end{subequations}
	 	
	 	\vspace{-1mm}
	 	\noindent and $\boldsymbol{V}_{\gamma_{S_l}} \triangleq  \left[\mathbf{v}_{S_1}^T,  \; \frac{P_P}{P_S}  \mathbf{v}_{S_2}^T,  \; \frac{P_P}{P_S} \mathbf{v}_{S_3}^T,  \; \frac{1}{\sqrt{P_S}} \right]$, where 
	 	\vspace{-1mm}
	 	\begin{subequations}	
	 		\begin{eqnarray}\label{eqn:v2}
	 		\!\!\!\!\!\!\!\!\!\!\!\!\!\!\!\!\!\mathbf{v}_{S_1} &\triangleq& \left[ \delta_{S_{11}} \beta_{S_{11}} \sqrt{\rho_{g_{11}} \zeta_{{g}_{1l}}} , \cdots , \delta_{S_{NL}} \beta_{S_{NL}} \sqrt{\rho_{g_{NL}} \zeta_{{g}_{Nl}}}  \, \right]^T\!\!\!, \\
	 		\!\!\!\!\!\!\!\!\!\!\!\!\!\!\!\!\!\mathbf{v}_{S_2} &\triangleq& \left[ \delta_{P_{11}} \beta_{P_{11}} \sqrt{\rho_{f_{11}} \zeta_{{v}_{1l}}} , \cdots , \delta_{P_{MK}} \beta_{P_{MK}} \sqrt{\rho_{f_{MK}} \zeta_{{v}_{Ml}}}  \, \right]^T\!\!\!, \\
	 		\!\!\!\!\!\!\!\!\!\!\!\!\!\!\!\!\!\mathbf{v}_{S_3} &\triangleq& \left[ \delta_{P_{1l}} \beta_{P_{1l}} c_{S_{1l}} \zeta_{{v}_{1l}} , \cdots , \delta_{P_{Ml}} \beta_{P_{Ml}} c_{S_{Ml}} \zeta_{{v}_{Ml}}  \, \right]^T\!\!\!.
	 		\end{eqnarray} 
	 	\end{subequations}
		
		\STATE If the status of the problem is feasible, then set $\lambda_{min} = \lambda_{\gamma}$, otherwise set $\lambda_{max} = \lambda_{\gamma}$.
		
		\STATE Stop if $\lambda_{max}-\lambda_{min} < \epsilon$. Otherwise go to Step 2.
		\ENDWHILE
		\RETURN $\eta_{P_{mk}} = \beta_{P_{mk}}^2 $ and $\eta_{S_{nl}} = \beta_{S_{nl}}^2 $for $m \in \{1, \cdots, M\}$, $k \in \{1, \cdots, K\}$,  $n \in \{1, \cdots, N\}$, and $l \in \{1, \cdots, L\}$.
	\end{algorithmic} 
\end{algorithm}

\section{The implication of DL pilot transmission}\label{sec:DL_pilot_tx}
In our achievable rate analysis in Section \ref{sec:Achievable sum rate analysis for primary system},   the PUs/SUs are assumed to be  unaware of DL channel estimates, and this  is a typical assumption in co-located massive MIMO  literature \cite{Marzetta2016_Book}. It is aimed at minimizing the pilot overhead to preserve system scalability. Thus,  the users  adopt  long-term  DL channel statistics  to decode  the received signals as the  DL channel coefficients  can be tightly approximated by their average counterparts by virtue of   channel hardening \cite{Ngo2017n}. Nevertheless, it has been shown in  \cite{Interdonato2016} that channel hardening in cell-free massive MIMO occurs only when a   large number of APs is distributed in close-vicinity, and hence, the adopting statistical DL CSI for signal decoding at users may hinder the system performance. 
To circumvent this,  DL pilots can be beamformed to estimate DL channels at the users, and this approach ensures that the DL pilot sequence length does not scale with the number of APs.   Next, we investigate the impact of DL pilots for the proposed cell-free massive MIMO with underlay spectrum sharing.

\subsection{DL channel estimation}\label{sec:DL_channel_estimation}
We define the effective DL desired and interference channel coefficients of the primary system based  on the signal received at  $U_P(k)$ in \eqref{eqn:k_PU_Rx_signal_reagd} as follows: 
\vspace{-1mm}
\begin{eqnarray} 
	\mu_{P_{ki}} \triangleq \sum\nolimits_{m=1}^{M} \delta_{P_{mi}} \eta_{P_{mi}}^{1/2} f_{mk} \hat{f}_{mi}^*  \qquad \text{and} \qquad
	\lambda_{P_{kj}} \triangleq \sum\nolimits_{n=1}^{N} \delta_{S_{nj}} \eta_{S_{nj}}^{1/2} u_{nk} \hat{g}_{nj}^*. \label{eqn:P_intrf_channel}
\end{eqnarray}

\vspace{-1mm}
\noindent To estimate DL channels, we need an additional $\tau_{p,d}$ symbol duration for transmitting DL pilots towards PUs/SUs. We consider the same pilot sharing technique that was used for  UL pilot transmission. The same pilot sequences will be used between P-APs/S-APs and PUs/SUs. For the sake of exposition, we denote the primary and secondary DL pilot sequences by $\boldsymbol {\phi}_{P_k,d}$, respectively, for $k \in \{1, \cdots, K\}$ and $\boldsymbol {\phi}_{S_l,d}$ for $l \in \{1, \cdots, L\}$. The pilot signal sent by the $m$th P-AP can be written as
\vspace{-1mm}
\begin{eqnarray}
	\mathbf{x}_{P_m,d}  &=&  \sqrt{P_{p,d}}   \sum\nolimits_{i=1}^{K}  \delta_{P_{mi}} \eta_{P_{mi}}^{1/2} \hat {f}_{mi}^* \boldsymbol {\phi}_{P_i,d} , \label{eqn:mth_tx_DL_signal} 
\end{eqnarray}

\vspace{-1mm}
\noindent  where $P_{p,d} \triangleq \tau_{p,d} P_{d}$ and $P_{d}$ is the average DL pilot transmit power at each P-AP/S-AP. 
Moreover, $\hat {f}_{mi}$  is defined in \eqref{eqn:estimate_f_mk}. Next, the  pilot vector received at  $U_P(k)$ can be written as
\vspace{-1mm}
\begin{eqnarray}
 	\mathbf y'_{P_k,d} &=& \sum\nolimits_{m=1}^{M} f_{mk} \mathbf{x}_{P_m,d} + \sum\nolimits_{n=1}^{N} u_{nk} \mathbf{x}_{S_n,d} + \mathbf n'_{P_k,d}, \label{eqn:P_rx_DL_pilot_signal}
\end{eqnarray}

\vspace{-1mm}
\noindent where $\mathbf n'_{P_k,d} $ is the AWGN   vector with i.i.d. $\mathcal {CN}(0,1)$ elements, at  $U_P(k)$. Then, we rewrite the received pilot vector at $U_P(k)$ by substituting \eqref{eqn:mth_tx_DL_signal}  into \eqref{eqn:P_rx_DL_pilot_signal}   as
\vspace{-1mm}
\begin{eqnarray}
	\mathbf y'_{P_k,d}  &=&  \sqrt{P_{p,d}} \left( \sum\nolimits_{i=1}^{K}  \mu_{P_{ki}}\boldsymbol {\phi}_{P_i,d}  +  \sum\nolimits_{j=1}^{L}  \lambda_{P_{kj}} \boldsymbol {\phi}_{S_j,d} \right) +  \mathbf n'_{P_k,d}, \label{eqn:P_rx_DL_pilot_signal_reang}
\end{eqnarray}

\vspace{-1mm}
\noindent where $\mu_{P_{ki}}$ and $\lambda_{P_{kj}}$ are the effective desired and interference DL channels   \eqref{eqn:P_intrf_channel}.
To estimate the  effective primary desired  DL channel, the sufficient statistics can be obtained by projecting  $ \boldsymbol {\phi}_{P_k,d}^H$  into \eqref{eqn:P_rx_DL_pilot_signal_reang}  as
\vspace{-2mm}
\begin{eqnarray}
	y_{P_k,d} &=& \boldsymbol {\phi}_{P_k,d}^H \mathbf y'_{P_k,d} = \sqrt{P_{p,d}} \left( \mu_{P_{kk}} +   \lambda_{P_{kk}}  \right) +  n_{P_k,d}, \label{eqn:P_projctd}
\end{eqnarray}

\vspace{-0mm}
\noindent where $n_{P_k,d} = \boldsymbol {\phi}_{P_k,d}^H \mathbf n'_{P_k,d}  \sim \mathcal{CN}(0,1)$ is the AWGN at  $U_P(k)$. 

\vspace{8mm}
\noindent \textit{\textbf{Proposition 3:}}
\textit{With beamformed DL pilots, the MMSE estimate  of $\mu_{P_{kk}}$ (\ref{eqn:P_intrf_channel}) is given by }
\vspace{-1mm}
\begin{eqnarray} \label{eqn:a_pkk_hat}
	\!\!\!\!\!\hat{\mu}_{P_{kk}} \! &=& \! \E{\mu_{P_{kk}}} \! +\! \frac{\Cov{\mu_{P_{kk}} y_{P_k,d}^*}}{\Cov{y_{P_k,d} y_{P_k,d}^*}} \left(y_{P_k,d} - \E{y_{P_k,d}}\right) \nonumber\\
	\!\!\!\!\!&=&  \E{\mu_{P_{kk}}} + \frac{ \sqrt{P_{p,d}} \Var{\mu_{P_{kk}}} }{ {P_{p,d}} \left(\Var{\mu_{P_{kk}}} + \Var{\lambda_{P_{kk}}}\right) + 1 }  \times \left(y_{P_k,d} - \sqrt{P_{p,d}} \left( \E{\mu_{P_{kk}}} + \E{\lambda_{P_{kk}}} \right)\right).
\end{eqnarray} 

\vspace{-1mm}

\noindent  \textit{Then,  the MMSE estimate of ${\mu}_{P_{kk}}$ is given  by evaluating   \eqref{eqn:a_pkk_hat} as}
\vspace{-1mm}
\begin{eqnarray}\label{eqn:a_pkk_hat_anly}
	\!\!\!\!\!\!\!\!\!\hat{\mu}_{P_{kk}}  \!&=&\! \frac{\!\sqrt{\!P_{p,d}} v_{P_{kk}} y_{P_k,d} \!+\!\! \sum\nolimits_{m=1}^{M} \!\delta_{P_{mk}} \eta_{P_{mk}}^{1/2} \rho_{f_{mk}} \!\!+\!  P_{p,d} \!\left(\! \!u_{P_{kk}} \!\!\sum\nolimits_{m=1}^{M} \!\delta_{P_{mk}} \eta_{P_{mk}}^{1/2} \rho_{f_{mk}} \!\!-\! v_{P_{kk}}\! \sum\nolimits_{n=1}^{N} \!\delta_{S_{nk}} \eta_{S_{nk}}^{1/2} \rho_{u_{nk}} \!  \right)\! }{P_{p,d} \left(v_{P_{kk}} + u_{P_{kk}}\right) + 1}.
\end{eqnarray}

\vspace{-1mm}
\noindent  \textit{In \eqref{eqn:a_pkk_hat_anly}, $v_{P_{kk}}$ and $u_{P_{kk}}$ are defined as}
\vspace{-1mm}
\begin{eqnarray} \label{eqn:v_pkk_&_u_pkk}
	v_{P_{kk}} \triangleq \sum\nolimits_{m=1}^{M} \delta_{P_{mk}} \eta_{P_{mk}} \zeta_{{f}_{mk}} \rho_{f_{mk}} \qquad \text{and} \qquad
	u_{P_{kk}} \triangleq \sum\nolimits_{n=1}^{N} \delta_{S_{nk}} \eta_{S_{nk}} \zeta_{{u}_{nk}} \rho_{u_{nk}}.
\end{eqnarray}

\vspace{-1mm}
\noindent \textit{The actual effective/desired DL channel coefficient is given by $\mu_{P_{kk}} = \hat{\mu}_{P_{kk}} + \epsilon^\mu_{P_{kk}}$, where $\epsilon^\mu_{P_{kk}}$ is estimation error, which is independent of respective channel estimate.}

\begin{proof}
	Appendix \ref{app:Appendix1_DL}. 
\end{proof}

\noindent  \textbf{\textit{Remark} 6}:
It is worth noting that  the  parameters  $\mu_{S_{lj}}$, $\lambda_{S_{li}}$, $\hat{\mu}_{S_{ll}}$, $v_{S_{ll}}$, and $u_{S_{ll}}$ corresponding to the secondary  system can be obtained by replacing subscripts $\{P,m,M, k, i\}$ in \eqref{eqn:P_intrf_channel}, \eqref{eqn:a_pkk_hat_anly}, and \eqref{eqn:v_pkk_&_u_pkk}, respectively, by  $\{S, n,N, l, j\}$.

\subsection{Primary achievable DL rate with DL pilots}\label{sec:P_DL_rate}
We can rewrite the  received signal at  $U_P(k)$ in \eqref{eqn:k_PU_Rx_signal_reagd} via the effective desired and interference DL channel coefficients as follows:
\vspace{-1mm}
\begin{eqnarray}\label{eqn:k_PU_DL_Rx_signal} 
	{r}_{P_k,d} &=&  \sqrt{P_{P}} \mu_{P_{kk}} q_{P_k} +  \sqrt{P_{P}} \sum\nolimits_{i \neq k}^{K} \mu_{P_{ki}} q_{P_i} +  \sqrt{P_S} \sum\nolimits_{j=1}^{L} \lambda_{P_{kj}} q_{S_j} + n_{P_k,d},
\end{eqnarray} 

\vspace{-1mm}
\noindent  where $n_{P_k,d} \sim \mathcal{CN}(0,1)$ is the AWGN at  $U_P(k)$. Then, by using the DL channel estimate at $U_P(k)$, \eqref{eqn:k_PU_DL_Rx_signal} can be rearranged to facilitate  decoding the desired signal as 
\vspace{-1mm}
\begin{eqnarray}\label{eqn:k_WC_Gaussian_Rx_signal_DL}
	{r}_{P_k,d} &=& \underbrace{\sqrt{P_{P}}    \E {\left( \mu_{P_{kk}} |\hat{\mu}_{P_{kk}} \right)  }q_{P_k} }_{\text{Desired signal}} + \underbrace{ \sqrt{P_{P} } \left(\left(\mu_{P_{kk}} |\hat{\mu}_{P_{kk}} \right) q_{P_k}  - \E{\left(\mu_{P_{kk}} |\hat{\mu}_{P_{kk}} \right)  } q_{P_k} \right) }_{\text{Detection uncertainty}} 
	\nonumber \\
	&&+ \underbrace{ \sqrt{P_P} \sum\nolimits_{i \neq k}^{K} \left(\mu_{P_{ki}} |\hat{\mu}_{P_{kk}} \right) q_{P_i} }_{\text{Inter-system interference}} + \underbrace{ \sqrt{P_S} \sum\nolimits_{j =1}^{L} \left(\lambda_{P_{kj}} |\hat{\mu}_{P_{kk}} \right) q_{S_j} }_{\text{Intra-system interference}}  + \underbrace{n_{P_k,d}}_{\text{AWGN}}.
\end{eqnarray}

\vspace{-1mm}
\noindent By using \eqref{eqn:k_WC_Gaussian_Rx_signal_DL}, the SINR at  $U_P(k)$ can be given as
\vspace{-1mm}
\begin{eqnarray}\label{eqn:P_SINR_DL}
	 \gamma_{P_k,d} = \frac{P_P  |\E{\mu_{P_{kk}} | \hat{\mu}_{P_{kk}}} |^2 }{P_P \sum\nolimits_{i=1}^{K} \E{| \mu_{P_{ki}} |^2 | \hat{\mu}_{P_{kk}} } - P_P |\E{\mu_{P_{kk}} | \hat{\mu}_{P_{kk}}} |^2 + P_S  \sum\nolimits_{j=1}^{L}  \E{|\lambda_{P_{kj}} |^2 | \hat{\mu}_{P_{kk}} }  + 1 }.
\end{eqnarray}	 

\vspace{-1mm}
\noindent  From the facts that (i) $\mu_{P_{kk}}$ is Gaussian distributed, (ii) $\hat{\mu}_{P_{kk}}$ and $\epsilon^\mu_{P_{kk}}$ are independent, and (iii) $\mu_{P_{kk}}$, $\mu_{P_{ki}}$,  and $\lambda_{P_{kj}}$ are independent for $i,j \neq k$ \cite{Interdonato2016}, we rewrite the SINR in \eqref{eqn:P_SINR_DL}  as follows:
\vspace{-1mm}
\begin{eqnarray}\label{eqn:P_SINR_DL_rearngd}
	\gamma_{P_k,d} &=&  \frac{P_P   | \hat{\mu}_{P_{kk}} |^2 }{P_P  \left( \sum\nolimits_{i\neq k}^{K}  \E{|\mu_{P_{ki}}|^2  }  +  \E{|\epsilon^\mu_{P_{kk}}|^2 } \right)  + P_S  \sum\nolimits_{j=1}^{L}  \E{|\lambda_{P_{kj}}|^2 } +  1 } .
\end{eqnarray}	 

\vspace{-1mm}
\noindent Thereby, we define the achievable DL rate as 
\vspace{-2mm}
\begin{eqnarray}\label{eqn:P_rate_DL}
	R_{P_k,d} =  \left({\tau_{d,d}}/{\tau_{c}}\right)\E[\hat{\mu}_{P_{kk}}]{\log[2] {1+ \gamma_{P_k,d}} },
\end{eqnarray}

\vspace{-2mm}
\noindent where $\tau_{d,d} \!=\! \tau_{c} \!-\! \left(\!\tau_{p} \!+\! \tau_{p,d} \!\right)$. From Jensen's inequality,  an upper bound for  DL rate at  $U_P(k)$ is derived as
\vspace{-2mm}
\begin{eqnarray}\label{eqn:P_rate_DL_ub}
	R^{ub}_{P_k,d} =  \left({\tau_{d,d}}/{\tau_{c}}\right)\log[2]{ 1+ \E[\hat{\mu}_{P_{kk}}]{\gamma_{P_k,d}} },
\end{eqnarray}

\vspace{-2mm}
\noindent where $\E[\hat{\mu}_{kk}]{\gamma_{P_k,d}}$  is given by
\vspace{-1mm}
\begin{eqnarray}\label{eqn:P_SINR_DL_rearngd_E}
	\E[\hat{\mu}_{P_{kk}}]{\gamma_{P_k,d}}  =  \frac{P_P  \E{ | \hat{\mu}_{P_{kk}} |^2} }{P_P   \left( \sum\nolimits_{i\neq k}^{K}  \E{|\mu_{P_{ki}}  |^2  }  +  \E{ |\epsilon^\mu_{P_{kk}}  |^2 }  \right)  +  P_S  \sum\nolimits_{j=1}^{L}  \E{|\lambda_{P_{kj}} |^2 }  +  1 }.
\end{eqnarray}	 

\vspace{-1mm}

\noindent \textit{\textbf{Proposition 4:}}
\noindent \textit{By evaluating expectation terms in \eqref{eqn:P_SINR_DL_rearngd_E},  the effective SINR at  $U_P(k)$  in the case of beamformed DL pilots by the P-APs can be derived as  follows:}
\vspace{-1mm}
\begin{eqnarray}\label{eqn:P_SINR_DL_rearngd_E_anly}
	\!\!\!\! \E[\hat{\mu}_{P_{kk}}]{\gamma_{P_k,d}}  \!=\! \frac{P_P \left( \sum\nolimits_{m=1}^{M} \sum\nolimits_{m'=1}^{M} \delta_{P_{mk}} \eta_{P_{mk}}^{1/2}  \delta_{P_{m'k}} \eta_{P_{m'k}}^{1/2} \rho_{f_{mk}} \rho_{f_{m'k}} + v_{P_{kk}} - \kappa^{\epsilon}_{P_{kk}} \right) }{\!P_P \! \left( \sum\nolimits_{i\neq k}^{K}  v_{P_{ki}} \!+\! \kappa^{\epsilon}_{P_{kk}}  \right) \!+\! P_S \!\left(  \sum\nolimits_{n=1}^{N}  \sum\nolimits_{j=1}^{L} \delta_{S_{nj}} \eta_{S_{nj}} \rho_{g_{nj}} \zeta_{{u}_{nk}} \!+\! \sum\nolimits_{n=1}^{N} \delta_{S_{nk}} \eta_{S_{nk}} \rho_{u_{nk}}^2  \right) \!+\! 1 },
\end{eqnarray}	 

\vspace{-1mm}
\noindent \textit{where $\kappa^{\epsilon}_{P_{kk}}$ and $v_{P_{ki}}$ are defined as}	
\vspace{-1mm}
\begin{eqnarray} \label{eqn:kappa_v_Pki} 
	\!\!\!\!\! \kappa^{\epsilon}_{P_{kk}} \!\triangleq\!  \frac{ \left(1 +  P_{p,d} u_{P_{kk}}\right)^2 v_{P_{kk}}  +  \left(P_{p,d} v_{P_{kk}}\right)^2 u_{P_{kk}}  +  P_{p,d} v_{P_{kk}}^2  }{ \left(P_{p,d} \left(v_{P_{kk}} + u_{P_{kk}}\right) +1 \right)^2} \quad \text{and} \quad
	v_{P_{ki}} \!\triangleq\! \sum\nolimits_{m=1}^{M} \delta_{P_{mi}} \eta_{P_{mi}} \rho_{f_{mi}} \zeta_{{f}_{mk}},
\end{eqnarray} 

\vspace{-1mm}
\noindent \textit{where $P_{p,d} \triangleq \tau_{p,d} P_{d}$}.
\begin{proof}
	  Appendix \ref{app:Appendix4}.
\end{proof}

\subsection{Secondary achievable DL rate with DL pilots}\label{sec:S_DL_rate}
We define an upper bound of the DL achievable rate at $U_S(l)$  via steps similar to \eqref{eqn:k_PU_DL_Rx_signal}-\eqref{eqn:P_rate_DL_ub}  as
\vspace{-2mm}
\begin{eqnarray}\label{eqn:S_rate_DL_ub}
	R^{ub}_{S_l,d} =  \left({\tau_{d,d}}/{\tau_{c}}\right)\log[2]{ 1+ \E[\hat{\mu}_{S_{ll}}]{\gamma_{S_l,d}} },
\end{eqnarray}

\vspace{-2mm}
\noindent where $\E[\hat{\mu}_{S_{ll}}]{\gamma_{S_l,d}}$  is given by
\vspace{-1mm}
\begin{eqnarray}\label{eqn:S_SINR_DL_rearngd_E}
	\E[\hat{\mu}_{S_{ll}}]{\gamma_{S_l,d}}  &=&   \frac{P_S  \E{ | \hat{\mu}_{S_{ll}} |^2 } }{P_S  \left(  \sum\nolimits_{j\neq l}^{L}  \E{|\mu_{S_{lj}}  |^2  }  +  \E{ |\epsilon^\mu_{S_{ll}}  |^2 }  \right)  + P_P  \sum\nolimits_{i=1}^{K}  \E{|\lambda_{S_{li}} |^2 }  +  1 }.
\end{eqnarray}	 

\vspace{-1mm}
\noindent Then, by following  steps similar to \eqref{eqn:P_SINR_DL_rearngd_E_anly} to  evaluate the expectation terms in  \eqref{eqn:S_SINR_DL_rearngd_E}, the effective SINR at $U_S(l)$ can be written as
\vspace{-1mm}
\begin{eqnarray}\label{eqn:S_SINR_DL_rearngd_E_anly}
	\!\!\!\!\!\!\ 	\!\!\!\!\! \E[\hat{\mu}_{S_{ll}}]{\gamma_{S_l,d}}  \!=\! \frac{P_S \left( \sum\nolimits_{n=1}^{N} \sum\nolimits_{n'=1}^{N}  \delta_{S_{nl}} \eta_{S_{nl}}^{1/2} \delta_{S_{n'l}} \eta_{S_{n'l}}^{1/2} \rho_{g_{nl}} \rho_{g_{n'l}} + v_{S_{ll}} - \kappa^{\epsilon}_{S_{ll}} \right) }{P_S  \left( \sum\nolimits_{j\neq l}^{L}  v_{S_{lj}} \!+\! \kappa^{\epsilon}_{S_{ll}}  \right) \!+\! P_P \left(  \sum\nolimits_{m=1}^{M}  \sum\nolimits_{i=1}^{K} \delta_{P_{mi}} \eta_{P_{mi}} \rho_{f_{mi}} \zeta_{{v}_{ml}} \!+\! \sum\nolimits_{m=1}^{M} \delta_{P_{ml}} \eta_{P_{ml}} \rho_{v_{ml}}^2  \right) \!+\!  1 },
\end{eqnarray}	 

\vspace{-1mm}
\noindent where $\kappa^{\epsilon}_{S_{ll}}$ and $v_{S_{lj}}$ are defined as
\vspace{-1mm}
\begin{eqnarray}\label{eqn:kappa_s_&_v_slj}
	\kappa^{\epsilon}_{S_{ll}} \triangleq  \frac{ \left(1+ P_{p,d} u_{S_{ll}}\right)^2 v_{S_{ll}} + \left(P_{p,d} v_{S_{ll}}\right)^2 u_{S_{ll}} + P_{p,d} v_{S_{ll}}^2  }{ \left(P_{p,d} \left(v_{S_{ll}} + u_{S_{ll}}\right) +1 \right)^2} \quad \text{and} \quad
	v_{S_{lj}} \triangleq \sum\nolimits_{n=1}^{N} \delta_{S_{nj}} \eta_{S_{nj}} \rho_{g_{nj}} \zeta_{{g}_{nl}}.
\end{eqnarray}	

\noindent  \textbf{\textit{Remark} 7}: We reveal through our numerical results in Section \ref{sec:Numerical}  that the adoption of  estimated DL CSI at the PUs/SUs  for signal decoding can be exploited to boost the achievable rates by circumventing the less prevalent channel hardening property in   cell-free massive MIMO compared to that of co-located counterpart. The underlying implication  is that the assumption of statistical DL channels are approximately equal to the instantaneous counterparts may not be accurate for cell-free massive MIMO.

\section{Achievable Rate analysis of NOMA-Aided Underlay Spectrum Sharing}\label{sec:NOMA underlay}
\subsection{Primary achievable rate with NOMA}\label{sec:noma_ach_rate}
We rearrange the received signal at $U_P(a,k)$   \eqref{eqn:noma_PRx_signal_aftr_sic}  to decode the desired signal as
\vspace{-1mm}
\begin{eqnarray}\label{eqn:noma_PRx_signal_aftr_sic_WCG}
	{r}_{P_{ak}} &=&   \underbrace{\sqrt{P_{P}}  \E{\sum\nolimits_{m=1}^{M}  \eta_{P_{mak}}^{1/2} f_{mak} \hat{f}_{mak}^* } q_{P_{ak}}}_{\text{Desired signal}} \nonumber \\
	&&+ \underbrace{ \sqrt{P_{P}} \sum\nolimits_{m=1}^{M}  \eta_{P_{mak}}^{1/2} f_{mak} \hat{f}_{mak}^*q_{P_{ak}}  - \sqrt{P_{P}} \E{\sum\nolimits_{m=1}^{M}  \eta_{P_{mak}}^{1/2} f_{mak} \hat{f}_{mak}^* } q_{P_{ak}}  }_{\text{Detection uncertinity}}   + w_{{P_{ak}}},
\end{eqnarray}

\vspace{-1mm}
\noindent where $w_{{P_{ak}}}$ denotes the effective noise at $U_P(a,k)$ containing   intra-cluster interference after SIC,  error propagation due to imperfect SIC,   intra-system interference,   inter-system interference, and   AWGN given in (\ref{eqn:noma_PRx_signal_aftr_sic}). From \eqref{eqn:noma_PRx_signal_aftr_sic_WCG}, we  write the SINR at $U_P(a,k)$  as
\vspace{-1mm}
\begin{eqnarray}\label{eqn:noma_SINR_ k_PU}
	\gamma_{P_{ak}}  =  \frac{{P_{P}} \left| \E {\sum\nolimits_{m=1}^{M} {\eta_{P_{mak}}^{1/2} f_{mak} \hat{f}_{mak}^*}} \right|^2 }{P_{P}\Var{\sum\nolimits_{m=1}^{N} {\eta_{P_{mak}}^{1/2} f_{mak} \hat{f}_{mak}^*}} + {P_{P}} \sum\nolimits_{i=1}^{4} {I_{P_{ai}}} +\E{\left|n_{P_{ak}}\right|^2} } ,
\end{eqnarray}

\vspace{-1mm}
\noindent where ${I_{P_{ai}}}$ for $i \in \{1,2,3,4\}$ can be defined as
\vspace{-1mm}
\begin{subequations}
	\begin{eqnarray}
	{I_{P_{a1}}} &=&  \E{\!\left|\sum\nolimits_{m=1}^{M} \sum\nolimits_{i=1}^{k-1} \eta_{P_{mai}}^{1/2} f_{mak} \hat{f}_{mak}^*  \right|^{2} },{\label{eqn:Ipa1} }  \\
	{I_{P_{a2}}} &=&  \E {\left| \sum\nolimits_{m=1}^{M} \sum\nolimits_{i=k+1}^{K} \left( f_{mak} \hat{f}_{mai}^* q_{P_{ai}} - \E{f_{mak} \hat{f}_{mai}^* } \hat{q}_{ai}  \right) \right|^{2} },{\label{eqn:Ipa2} } \\
	{I_{P_{a3}}} &=&  \E {\left|\sum\nolimits_{m=1}^{M} \sum\nolimits_{a' \neq a}^{A} \sum\nolimits_{i=1}^{K} \eta_{P_{m a' i}}^{1/2} f_{mak} \hat{f}_{m a' i}^*  \right|^2 }, \nonumber \\ 
	{I_{P_{a4}}} &=& \frac{{P_{S}}}{{P_{P}}} \E {\left|\sum\nolimits_{n=1}^{N} \sum\nolimits_{ b = 1}^{B} \sum\nolimits_{j=1}^{L} \eta_{S_{n b j}}^{1/2} u_{nak} \hat{g}_{n b j}^*  \right|^2 }. {\label{eqn:Ipa4} } 
	\end{eqnarray} 
\end{subequations}

\noindent  Then, we compute the SINR by evaluating the expectation and variance terms in \eqref{eqn:noma_SINR_ k_PU}  as 
\vspace{-1mm}
\begin{eqnarray}\label{eqn:noma_sinr_k_PU}
	\!\!\!\!\!\! \gamma_{P_{ak}} \!=\! \frac{P_{P} \left(\sum\nolimits_{m=1}^{M} \eta_{P_{mak}}^{1/2} \alpha_{f_{mak}} \right)^{2}} { P_{P} \! \sum\nolimits_{m=1}^{M}  \sum\nolimits_{a=1}^{A} \sum\nolimits_{i=1}^{K} \eta_{P_{ma i}} \alpha_{f_{ma i}} \zeta_{f_{mak}} \!+\! {P_{P}}\! \sum\nolimits_{i=1}^{3} I'_{P_{ai}} \!+\! \sum\nolimits_{n=1}^{N} \sum\nolimits_{b=1}^{B} \sum\nolimits_{j=1}^{L} \eta_{S_{nb j}} \alpha_{g_{nb j}} \zeta_{{u}_{nak}} \!+\! 1 },
\end{eqnarray}

\vspace{-1mm}
\noindent where $I'_{P_{ai}}$ for $i\in\{1,2,3\}$ is defined as
\vspace{-1mm}
\begin{subequations}
	\begin{eqnarray}
	{I'_{P_{a1}}} &=&  \sum\nolimits_{m=1}^{M} \sum\nolimits_{i=1}^{k-1}  \eta_{P_{mai}} \alpha_{f_{mak}}^2  \left( \frac{\zeta_{{f}_{mai}}}{ \zeta_{{f}_{mak}}} \right)^{2},{\label{eqn:Ipa1_dash} } \\ 
	{I'_{P_{a2}}} &=& 2  \sum\nolimits_{m=1}^{M} \sum\nolimits_{i=k+1}^{K}  (1-\vartheta_{P_{ai}})\eta_{P_{mai}} \alpha_{f_{mak}}^{2} \left(\frac{\zeta_{{f}_{mai}}}{ \zeta_{{f}_{mak}}}\right)^{2},   {\label{eqn:Ipa2_dash} } \\
	{I'_{P_{a3}}}  &=& \frac{{P_{S}}}{{P_{P}}} \sum\nolimits_{n=1}^{N} \sum\nolimits_{j=1}^{L} \eta_{S_{naj}} \alpha_{f_{mak}}^2 \left(\frac{\zeta_{{u}_{nak}} \zeta_{{g}_{naj}}}{\zeta_{{f}_{mak}} } \right)^{2}.{\label{eqn:Ipa3_dash} }  
	\end{eqnarray} 
\end{subequations}
The derivation of (\ref{eqn:noma_sinr_k_PU}) follows steps similar to those in Appendix \ref{app:Appendix3}, and hence, it is omitted for the sake of brevity.  
Next, the achievable rate of $U_P(a,k)$ and the   sum rate of  primary system are given by
\vspace{-1mm}
\begin{eqnarray}\label{eqn:noma_rate_of_k_PU}
	R_{P_{ak}} = \left({\left(\tau_c -\tau_p\right)}/{\tau_c}\right)\log[2] {1+ \gamma_{P_{ak}}}   \quad \text{and}\quad R_{P} = \sum\nolimits_{a=1}^{A} \sum\nolimits_{k=1}^{K} R_{P_{ak}},\label{eqn:noma_rate_P_sum}
\end{eqnarray}

\vspace{-1mm}
\noindent where $\gamma_{P_{ak}}$ is defined in (\ref{eqn:noma_sinr_k_PU}). 

\subsection{Secondary achievable rate with NOMA}\label{sec:noma_ach_rate_S}
We follow a similar analysis to Section \ref{sec:noma_ach_rate} for  deriving the   sum rate of the secondary system  as
\vspace{-1mm}
\begin{eqnarray}\label{eqn:noma_rate_S_sum}
	R_{S} = \sum\nolimits_{b=1}^{B} \sum\nolimits_{l=1}^{L} R_{S_{bl}},
\end{eqnarray}

\vspace{-1mm}
\noindent where $ R_{S_{bl}}$ is the achievable rate of $U_S(b,l)$ and   given by
\vspace{-2mm}
\begin{eqnarray}\label{eqn:noma_rate_of_l_SU}
	R_{S_{bl}} = \left({\left(\tau_c -\tau_p\right)}/{\tau_c}\right)\log[2] {1+ \gamma_{S_{bl}}}.
\end{eqnarray}

\vspace{-2mm}
\noindent In \eqref{eqn:noma_rate_of_l_SU}, we obtain the SINR at $U_S(b,l)$ denoted by $\gamma_{S_{bl}}$  by following (\ref{eqn:noma_sinr_k_PU}) and replacing the primary system variables with respective secondary system variables in  \eqref{eqn:noma_sinr_k_PU} as
\vspace{-1mm}
\begin{eqnarray}\label{eqn:noma_sinr_l_SU}
	\!\!\!\!\!\!\! \gamma_{S_{bl}} \!=\! \frac{P_{S} \left(\sum\nolimits_{n=1}^{N} \eta_{S_{nbl}}^{1/2} \alpha_{g_{nbl}} \right)^{2}} { P_{S} \sum\nolimits_{n=1}^{N}  \sum\nolimits_{b=1}^{B} \sum\nolimits_{j=1}^{L} \eta_{S_{nb j}} \alpha_{g_{nb j}} \zeta_{g_{nbl}} \!+\! P_{S} \sum\nolimits_{j=1}^{3} I'_{S_{bj}} \!+\! \sum\nolimits_{m=1}^{M} \sum\nolimits_{a=1}^{A} \sum\nolimits_{j=1}^{K} \eta_{P_{ma i}} \alpha_{f_{ma i}} \zeta_{{v}_{mbl}} \!+\!1 },
\end{eqnarray}

\vspace{-1mm}
\noindent where $I'_{S_{bi}}$ for $i\in\{1,2,3\}$ is given by
\vspace{-1mm}
\begin{subequations}
	\begin{eqnarray}
	{I'_{S_{b1}}} &=&  \sum\nolimits_{n=1}^{N} \sum\nolimits_{j=1}^{l-1} \eta_{S_{nbj}} \alpha_{g_{nbl}}^2  \left(\frac{\zeta_{{g}_{nbj}}}{ \zeta_{{g}_{nbl}}}\right)^{2} ,{\label{eqn:Isb1_dash} }  \\
	{I'_{S_{b2}}} &=& 2  \sum\nolimits_{n=1}^{N} \sum\nolimits_{j=l+1}^{L}  (1-\vartheta_{S_{bj}})\eta_{S_{nbj}} \alpha_{g_{nbl}}^2  \left(\frac{\zeta_{{g}_{nbj}}}{ \zeta_{{g}_{nbl}}}\right)^{2}  ,{\label{eqn:Isb2_dash} } \\
	{I'_{S_{b3}}} &=& \frac{{P_{P}}}{P_{S}} \sum\nolimits_{m=1}^{M} \sum\nolimits_{i=1}^{K} \eta_{P_{mbi}} \alpha_{g_{nbl}}^2 \left(\frac{\zeta_{{v}_{mbl}} \zeta_{{f}_{mbi}}}{\zeta_{{g}_{nbl}} } \right)^2. {\label{eqn:Isb3_dash} }  
	\end{eqnarray} 
\end{subequations}

\section{The implication of DL pilot transmission with NOMA}\label{sec:DL_pilot_NOMA}
The effective DL  desired and interference channel coefficients at $U_P(a,k)$ are defined from \eqref{eqn:noma_k_PU_Rx_signal} as
\vspace{-2mm}
\begin{eqnarray}
	\mu_{P_{a'i}^{ak}} \triangleq \sum\nolimits_{m=1}^{M} \eta_{P_{m a' i}}^{1/2} f_{mak} \hat{f}_{ma'i}^*  \qquad \text{and} \qquad  \lambda_{P_{bj}^{ak}} \triangleq \sum\nolimits_{n=1}^{N} \eta_{S_{nbj}}^{1/2} u_{nak} \hat{g}_{nbj}^*. \label{eqn:mu_pak_lambda_pak}   
\end{eqnarray} 

\vspace{-1mm}
\noindent The P-APs/S-APs again use the same pilot sequences   in \eqref{eqn:noma_pilot_sequences} to beamform DL pilots toward  PUs/SUs.  
Then, a sufficient statistic to estimate the desired effective DL channel at $U_P(a,k)$ is given by 
\vspace{-1mm}
\begin{eqnarray}
	y_{P_{ak},d} = \sqrt{P_{p,d}} \left( \sum\nolimits_{i=1}^{K} \mu_{P_{ai}^{ak}} +  \sum\nolimits_{j=1}^{L} \lambda_{P_{aj}^{ak}}  \right) +  n_{P_{ak},d}, \label{eqn:noma_P_projctd}
\end{eqnarray}

\vspace{-1mm}
\noindent where $n_{P_{ak},d}  \sim \mathcal{CN}(0,1)$ is the AWGN at $U_P(a,k)$.  
Thereby, the MMSE estimate  of $\mu_{P_{at}^{ak}}$ can be derived as \cite{Interdonato2016,Kay1993}  
\vspace{-1mm}
\begin{eqnarray} \label{eqn:mu_pak_hat}
	\!\!\!\!\!\hat{\mu}_{P_{at}^{ak}} &=& \E{\mu_{P_{at}^{ak}}} \! +\! \frac{\Cov{\mu_{P_{at}^{ak}} y_{P_{ak},d}^*}}{\Cov{y_{P_{ak},d} y_{P_{ak},d}^*}} \left(y_{P_{ak},d} - \E{y_{P_{ak},d}}\right) \nonumber\\
	\!\!\!\!\!&=&  \E{\mu_{P_{at}^{ak}}} + \frac{ \sqrt{P_{p,d}} \Var{\mu_{P_{at}^{ak}}} }{ {P_{p,d}} \left( \sum\nolimits_{i=1}^{K} \Var{\mu_{P_{ai}^{ak}}} + \sum\nolimits_{j=1}^{L} \Var{\lambda_{P_{aj}^{ak}}}\right) + 1 }  \times \left(y_{P_k,d} - \E{y_{P_{ak},d}}\right).
\end{eqnarray} 
 
\vspace{-1mm}
\noindent By evaluating \eqref{eqn:mu_pak_hat} via steps similar to those in Appendix \ref{app:Appendix1_DL}, the MMSE estimate of $\mu_{P_{at}^{ak}}$ can be derived as 
\vspace{-1mm} 
\begin{eqnarray} \label{eqn:mu_pak_hat_1}
	\hat{\mu}_{P_{at}^{ak}} =  \theta_{P_{at}^{ak}} + \Omega_{P_{at}^{ak}}\left(y_{P_k,d} - \sqrt{P_{p,d}} \left(\sum\nolimits_{i=1}^{K} \theta_{P_{ai}^{ak}} + \sum\nolimits_{j=1}^{L} \psi_{P_{aj}^{ak}} \right) \right),
\end{eqnarray}  

\vspace{-1mm}
\noindent where $\theta_{P_{ai}^{ak}} = \sum\nolimits_{m=1}^{M} \eta_{P_{mai}}^{1/2} \alpha_{f_{mak}} \zeta_{{f}_{mai}} /\zeta_{{f}_{mak}}$, $\psi_{P_{aj}^{ak}} = \sum\nolimits_{n=1}^{N} \eta_{S_{naj}}^{1/2} \alpha_{u_{nak}} \zeta_{{g}_{nak}} / \zeta_{{g}_{naj}}$, and 
\vspace{-1mm}
\begin{subequations}
	\begin{eqnarray}
	\!\!\!\!\Omega_{P_{at}^{ak}} &=& {\sqrt{P_{p,d}} \left( \varrho_{P_{at}^{ak}}^{\mu} - \theta_{P_{at}^{ak}}^2 \right)  } \Big/ \left({ P_{p,d} \left(  \sum\nolimits_{i=1}^{K} \left( \varrho_{P_{ai}^{ak}}^{\mu} - \theta_{P_{ai}^{ak}}^2 \right) + \sum\nolimits_{j=1}^{L}  \left( \varrho_{P_{bj}^{ak}}^{\lambda}  - \psi_{P_{aj}^{ak}}^2 \right) \right) + 1 }\right),\\
	\!\!\!\!\varrho_{P_{ai}^{ak}}^{\mu} &=& \sum\nolimits_{m=1}^{M}  \eta_{P_{ma i}} \alpha_{f_{ma i}} \left(\zeta_{{f}_{mak}}+ \alpha_{f_{mak}}\right), \label{eqn:varho_mu_pak} \quad \text{and}\quad
	\varrho_{P_{bj}^{ak}}^{\lambda} = \sum\nolimits_{n=1}^{N} \eta_{S_{nbj}}\left( \zeta_{{u}_{nak}} \alpha_{g_{nb j}} + \right).  \label{eqn:varho_lambda_pak} 
	\end{eqnarray} 
\end{subequations}

\vspace{-1mm}
\noindent The actual effective DL channel gain is given by ${\mu}_{P_{at}^{ak}} = \hat{\mu}_{P_{at}^{ak}} + \epsilon_{P_{at}^{ak}}^{\mu}$, where  $\epsilon_{P_{at}^{ak}}^{\mu}$ is an estimation error, which is  independent of the channel estimate $\hat{\mu}_{P_{at}^{ak}}$.

\subsection{Primary achievable DL rate with DL pilots in NOMA} \label{rate_DL_noma}
It is assumed that the users in  same cluster are ordered based on the effective channel gains as per  \eqref{eqn:noma_eff_chn_strngth}, and hence, SIC can be adopted to decode the power-domain NOMA  signals at the users  \cite{Ding2016}. Thereby, the post-processed signal after an imperfect SIC operation at $U_P(a,k)$ can be written as 
\vspace{-1mm}
\begin{eqnarray}\label{eqn:noma_DL_signal_aftr_sic}
	{r}_{P_{ak},d} &=& \underbrace{\sqrt{P_{P}}  \mu_{P_{ak}^{ak}} q_{P_{ak}} }_{\text{Desired signal}} + \underbrace{ \sqrt{P_P} \sum\nolimits_{i=1}^{k-1}  \mu_{P_{ai}^{ak}} q_{P_{ai}} }_{\text{Intra-cluster interference after SIC}} + \underbrace{ \sqrt{P_P} \sum\nolimits_{i=k+1}^{K} \epsilon_{P_{ai}^{ak}}^{\mu} q_{P_{ai}}  }_{\text{Error propagation due to imperfect SIC}} \nonumber \\
	&&+ \underbrace{ \sqrt{P_P} \sum\nolimits_{a' \neq a}^{A} \sum\nolimits_{i =1}^{K} \mu_{P_{a'i}^{ak}}  q_{P_{a' i}} }_{\text{Intra-system interference}} + \underbrace{ \sqrt{P_S} \sum\nolimits_{b =1}^{B} \sum\nolimits_{j =1}^{L} \lambda_{P_{bj}^{ak}} q_{S_{b j}} }_{\text{Inter-system interference}} + \underbrace{n_{P_{ak}}}_{\text{AWGN}}.
\end{eqnarray}

\vspace{-1mm}
\noindent By using \eqref{eqn:noma_DL_signal_aftr_sic}, the corresponding SINR at $U_P(a,k)$ can be derived as  
\vspace{-1mm}
\begin{eqnarray}\label{eqn:P_SINR_DL_noma}
	\!\!\!\!\!\!\!\! \gamma_{P_{ak},d} \!=\! \frac{P_P  \left|\E{\mu_{P_{ak}^{ak}} | \hat{\mu}_{P_{ak}^{ak}}} \right|^2 }{\!P_P \!\! \sum\nolimits_{a'=1}^{A} \!\sum\nolimits_{i=1}^{K} \! \E{\left| \mu_{P_{a'i}^{ak}} | \hat{\mu}_{P_{kk}}  \right|^{\!2}  } \!\!-\! P_P \!\! \sum\nolimits_{i=k}^{K} \!\left|\E{\mu_{P_{ak}^{ak}}| \hat{\mu}_{P_{ak}^{ak}}} \!\right|^{\!2} \!\!+\! P_S \!\! \sum\nolimits_{b=1}^{B} \! \sum\nolimits_{j=1}^{L} \! \E{\left|\lambda_{P_{aj}^{ak}} | \hat{\mu}_{P_{ak}^{ak}} \right|^{\!2}}  \!\!+\! 1 \!}\!.
\end{eqnarray}	

\vspace{-1mm}
\noindent By using  techniques similar to those used in \eqref{eqn:P_SINR_DL_rearngd}, the expected value of  $\gamma_{P_{ak},d}$ can be written as 
\vspace{-1mm}
\begin{eqnarray}\label{eqn:P_SINR_DL_rearngd_noma}
 	\E[\hat{\mu}_{P_{ak}^{ak}}]{\gamma_{P_{ak},d}} &=&  \frac{P_P   \E{\left| \hat{\mu}_{P_{ak}^{ak}} \right|^2} }{P_P   \sum\nolimits_{a'\neq a}^{A}\sum\nolimits_{i=k}^{K} \E{ \left| \mu_{P_{ai}^{ak}} \right|^2  }   +\sum_{j=1}^{2} I^{\hat{\mu}}_j  +P_S  \sum\nolimits_{b=1}^{B} \sum\nolimits_{j=1}^{L}  \E{\left|\lambda_{P_{bj}^{ak}} \right|^2 } +  1 },
\end{eqnarray}	 

\vspace{-1mm}
\noindent where $I^{\hat{\mu}}_j$ for $j \in \{1,2\}$ is given as
\begin{subequations}
\begin{eqnarray}\label{eqn:I_mu_def}
	I^{\hat{\mu}}_1 &=& P_P  \sum\nolimits_{i=k+1}^{K}  \E{\left|\hat{\mu}_{P_{ak}^{ak}} \right|^2}  \quad \text{and} \quad I^{\hat{\mu}}_2 =  P_P  \sum\nolimits_{i=1}^{K}  \E{\left|\epsilon^\mu_{P_{ai}^{ak}} \right|^2 }.
\end{eqnarray}
\end{subequations}

\vspace{-1mm}
\noindent By evaluating the expectation terms in \eqref{eqn:P_SINR_DL_rearngd_noma}, the average SINR at $U_P(a,k)$ can be  derived as 
\vspace{-1mm}
\begin{eqnarray}\label{eqn:P_SINR_DL_noma_ana}
	\!\!\!\!\!\!\!\!\! \E[\hat{\mu}_{P_{ak}^{ak}}]{\gamma_{P_{ak},d}} \!&=& \! \frac{P_P \Phi_{P_{ak}^{ak}}^{\mu}  }{\! P_P \! \sum\nolimits_{a'\neq a}^{A} \!\sum\nolimits_{i=1}^{K} \! \varrho_{P_{ai}^{ak}}^{\mu}  \!+\!  P_P \! \sum\nolimits_{i=k+1}^{K} \! \Phi_{P_{ak}^{ak}}^{\mu}  \!+\!  P_P  \! \sum\nolimits_{i=1}^{K} \!\!\left(\! \varrho_{P_{ai}^{ak}}^{\mu} \!-\! \Phi_{P_{ai}^{ak}}^{\mu} \!\right) \!\!+\! P_S\! \sum\nolimits_{b=1}^{B} \!\sum\nolimits_{j=1}^{L} \!\varrho_{P_{bj}^{ak}}^{\lambda}   \!\!+\! 1 \!},
\end{eqnarray}	 

\vspace{-1mm}
\noindent where  $\Phi_{P_{ai}^{ak}}^{\mu}$ is given by
\vspace{-1mm}
\begin{eqnarray}
	\!\!\!\!\!\Phi_{P_{ai}^{ak}}^{\mu} &\triangleq& \theta_{P_{at}^{ak}}^2 \!+\! \Omega_{P_{at}^{ak}}^2\left( {P_{p,d}} \left(\sum\nolimits_{i=1}^{K} \varrho_{P_{ai}^{ak}}^{\mu} \!+\! \sum\nolimits_{j=1}^{L} \varrho_{P_{bj}^{ak}}^{\lambda} \right)\!+\! 1 \!-\!  {P_{p,d}} \left(\sum\nolimits_{i=1}^{K} \theta_{P_{ai}^{ak}}\! +\! \sum\nolimits_{j=1}^{L} \psi_{P_{aj}^{ak}} \right)^2 \right).\label{eqn:Phi_pak} 
\end{eqnarray} 

\vspace{-1mm}
\noindent Then, an upper bound for the   DL rate at $U_P(a,k)$ with   estimated DL channels via the beamformed pilots can be derived as
$R^{ub}_{P_{ak},d} \!=\!  \left({\tau_{d,d}}/{\tau_{c}}\right)\log[2]{ \!1\!+\! \E[\hat{\mu}_{P_{ak}^{ak}}]{\gamma_{P_{ak},d}} }$,
where $\E[\hat{\mu}_{P_{ak}^{ak}}]{\gamma_{P_{ak},d}}$ is defined in \eqref{eqn:P_SINR_DL_noma_ana}.

\subsection{Secondary achievable DL rate with DL pilots in NOMA} \label{S_rate_DL_noma}
By following an analysis similar to  Section   \ref{rate_DL_noma}, the achievable  rate at $U_S(b,l)$  with   estimated DL channels  can be derived as
$R^{ub}_{S_{bl},d} =  \left({\tau_{d,d}}/{\tau_{c}}\right)\log[2]{ 1+ \E[\hat{\mu}_{P_{bl}^{bl}}]{\gamma_{S_{bl},d}} }$,
where $\E[\hat{\mu}_{P_{bl}^{bl}}]{\gamma_{S_{bl},d}}$ is defined as
\vspace{-1mm}
\begin{eqnarray}\label{eqn:S_SINR_DL_noma_ana}
	\!\!\!\!\!\!\!\!\! \E[\hat{\mu}_{S_{bl}^{bl}}]{\gamma_{S_{bl},d}} \!&=& \!  \frac{P_S \Phi_{S_{bl}^{bl}}^{\mu}  }{\! P_S \!\sum\nolimits_{b'\neq b}^{B}\sum\nolimits_{j=1}^{L} \varrho_{S_{bj}^{bl}}^{\mu} \! +\!  P_S\! \sum\nolimits_{j=l+1}^{L} \Phi_{S_{bl}^{bl}}^{\mu}  \!+\!  P_S  \! \sum\nolimits_{j=1}^{L} \!\left( \varrho_{S_{bj}^{bl}}^{\mu} \!-\! \Phi_{S_{bj}^{bl}}^{\mu} \right) \!+\! P_P \! \sum\nolimits_{a=1}^{A} \sum\nolimits_{i=1}^{K} \varrho_{S_{ai}^{bl}}^{\lambda}   \!+\! 1 },
\end{eqnarray}	

\vspace{-1mm}
\noindent where $\Phi_{S_{bl}^{bl}}^{\mu} $, $\varrho_{S_{bj}^{bl}}^{\mu} $, and $\varrho_{S_{ai}^{bl}}^{\lambda}$ can be obtained by replacing the primary subscripts $\{P, M, m, A, a, K, k, i\}$ with respective secondary subscripts $\{S, N, n, B, b, L, l, j\}$ in  \eqref{eqn:Phi_pak},  \eqref{eqn:varho_mu_pak}, and \eqref{eqn:varho_lambda_pak}, respectively.

\section{Numerical Results}\label{sec:Numerical}
In this section, our numerical results are presented to obtained useful insights. The simulation parameters are as follows: $\tau_{c} = 196$, $\tau_{p} = \tau_{p,d} = \max(K,L)$, and $\zeta_{{ h}_{ab}} = (d_0/d_{ab})^{\nu} \times 10^{\varphi_{ab/10}}$, where $d_{ab}$ is transmission distance between the $a$th P-AP/S-AP and the $b$th PU/SU, $d_0$ is the reference distance, and $\nu$ is the path-loss exponent. Here, $10^{\varphi_{ab/10}}$ captures the shadow fading with $\varphi_{ab} \sim \mathcal{N}(0,8)$. In an area of    $800 \times 800 \,m^2 $, the P-APs/S-APs are uniformly distributed, while PUs/SUs are randomly placed. 

\begin{figure}[!t]\centering\vspace{-7mm}
	\includegraphics[width=0.45\textwidth]{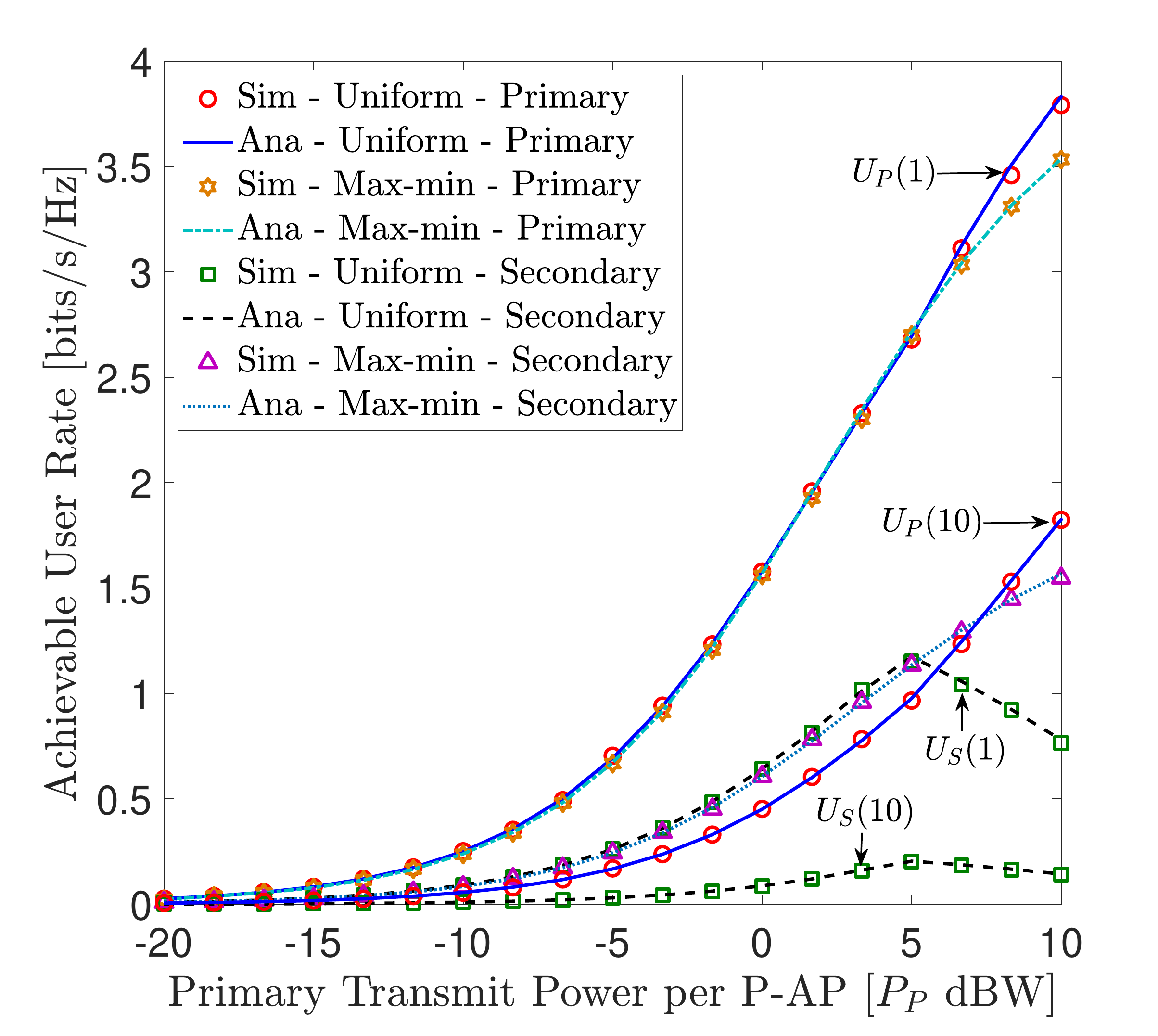}\vspace{-3mm}
	\caption{User rate versus the primary transmit power for $K = L = 10$, $M=N=32$, $I_T=0$ dB,  $\sigma_P^2 = \sigma_S^2 = 1$, $\nu = 2.4$, and $d_0 = 1$\,m. Moreover, $P_S=P_P/2$ and $P_{S_n} = \min{\left(P_{S}, {I_{T_1}}\big/{Z_{1}}, \cdots, {I_{T_k}}\big/{Z_{k}}, \cdots, {I_{T_K}}\big/{Z_{K}} \right)}$. }
	\label{fig:US_rate_vs_SNR_eql_max_min}\vspace{-10mm}
\end{figure}

In Fig. \ref{fig:US_rate_vs_SNR_eql_max_min}, the implications of our max-min based multi-objective transmit power allocation are investigated. To this end,  the achievable user rates of the primary and secondary systems are plotted against the primary transmit power per P-AP ($P_P$). The maximum allowable secondary transmit power ($P_{S}$) at each S-AP is kept at $P_P/2$. Then, the primary and secondary transmit power control coefficients are computed by using the proposed max-min algorithm in Section \ref{sec:Power Control}. The rates with max-min power allocation are compared with those with uniform power allocation. The pair \{$U_P(1), U_S(1)$\} is the users with strongest channels, while the pair \{$U_P(10), U_S(10)$\} represents the users with weakest channel gains. 
Fig. \ref{fig:US_rate_vs_SNR_eql_max_min} reveals that PUs/SUs experience distinct achievable rates when the uniform power allocation is adopted. Thus, the   achievable rates are dependent on the detrimental near-far effects. When the proposed max-min power control is employed, all  PUs/SUs achieve their respective  common rates regardless of the near-far effects.  For instance,  at $P_P=0$\,dBW, 
the weaker user $U_P(10)$ achieves a rate gain of $241.3\%$  from our max-min power allocation   over the uniform power allocation. 

\begin{figure}[!t]\centering\vspace{-7mm}
	\includegraphics[width=0.45\textwidth]{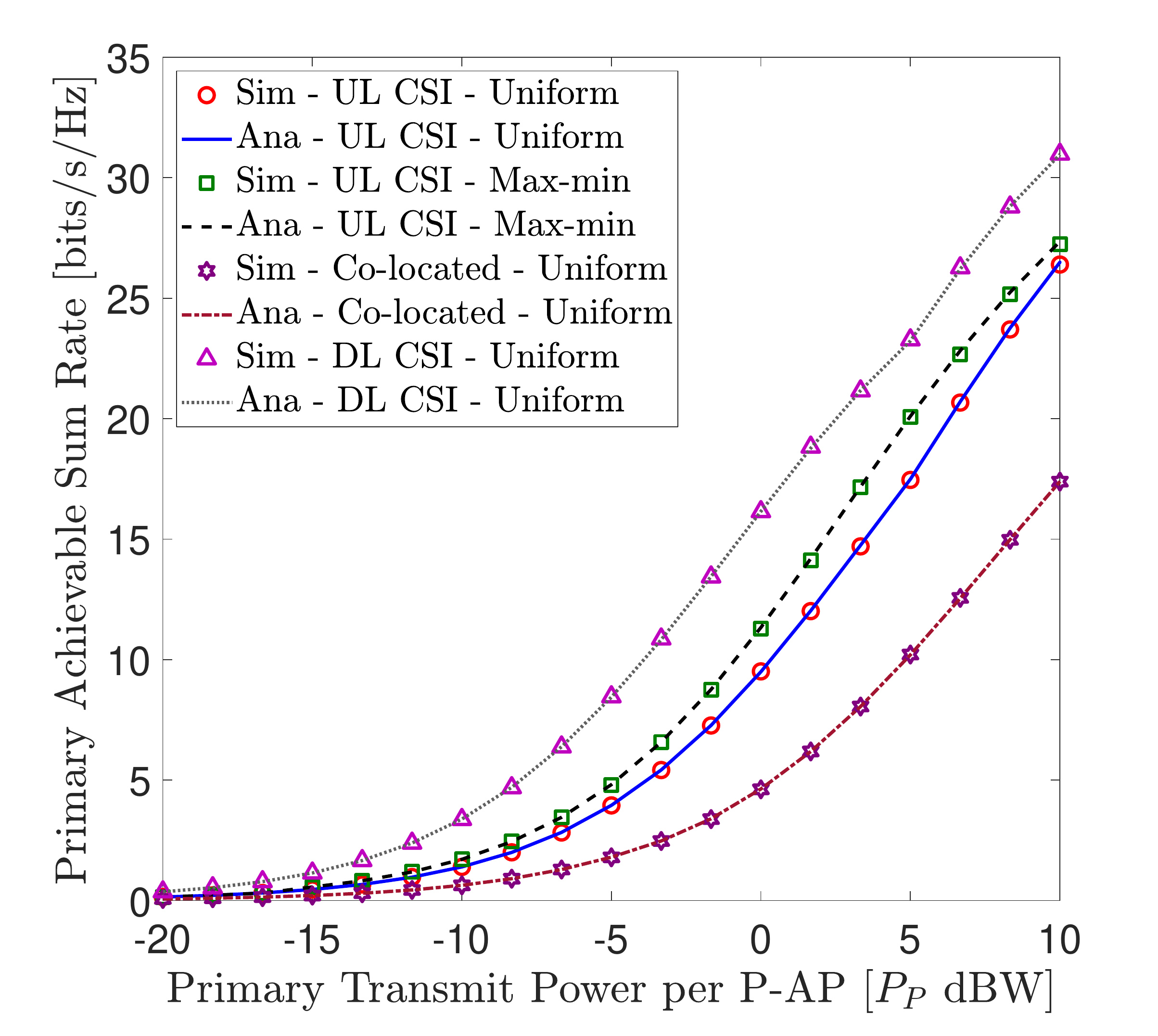}\vspace{-3mm}
	\caption{The primary  sum rate versus the primary transmit power for $K = L = 10$, $M=N=32$, $I_T = 0 \, dB$, $\sigma_P^2 = \sigma_S^2 = 1$, $\nu = 2.4$, and $d_0 = 1$\,m. }
	\label{fig:primary_sum_rate_vs_SNR_eql_max_min_co_loctd_DL_pilots}\vspace{-10mm}
\end{figure}

\begin{figure}[!t]\centering\vspace{-7mm}
	\includegraphics[width=0.45\textwidth]{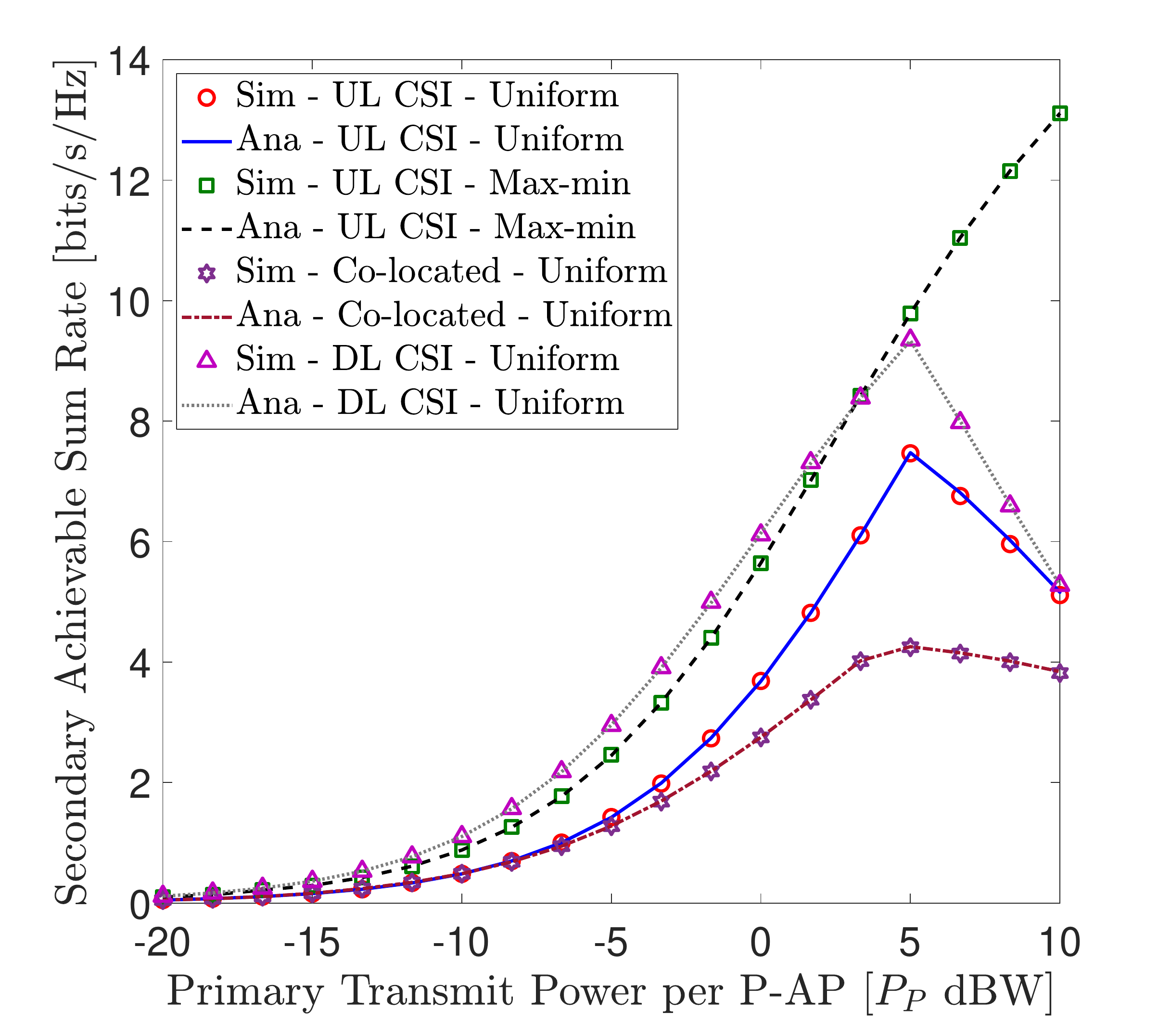}\vspace{-3mm}
	\caption{The secondary  sum rate versus the primary transmit power for $K = L = 10$, $M=N=32$, $I_T = 0 \, dB$, $\sigma_P^2 = \sigma_S^2 = 1$, $\nu = 2.4$, and $d_0 = 1$\,m. Moreover, $P_S=P_P/2$ and $P_{S_n} = \min{\left(P_{S}, {I_{T_1}}\big/{Z_{1}}, \cdots, {I_{T_k}}\big/{Z_{k}}, \cdots, {I_{T_K}}\big/{Z_{K}} \right)}$.   }
	\label{fig:secondary_sum_rate_vs_SNR_eql_max_min_co_loctd_DL_pilots}\vspace{-0mm}
\end{figure}

In Fig. \ref{fig:primary_sum_rate_vs_SNR_eql_max_min_co_loctd_DL_pilots} and Fig. \ref{fig:secondary_sum_rate_vs_SNR_eql_max_min_co_loctd_DL_pilots}, an achievable rate comparison is presented for cell-free/co-located, statistical/estimated DL CSI with max-min/uniform power allocations. In this context, the achievable sum rates of the primary and secondary systems, respectively, are plotted in Fig. \ref{fig:primary_sum_rate_vs_SNR_eql_max_min_co_loctd_DL_pilots} and Fig. \ref{fig:secondary_sum_rate_vs_SNR_eql_max_min_co_loctd_DL_pilots} as a function of the primary transmit power per P-AP ($P_P$).  In Fig. \ref{fig:secondary_sum_rate_vs_SNR_eql_max_min_co_loctd_DL_pilots}, $P_{S}$ is set to $P_P/2$ at each S-AP. When the uniform power allocation is adopted,  the achievable sum rate of the primary system increases monotonically with $P_P$.
However, for the secondary system, the achievable sum rate gradually increases up to a maximum in the low $P_P$ regime since $P_{S}$ is proportional to $P_P$, and then it decreases as $P_P$ grows without bound.
The reason for this behavior is that when $P_P$ increases,  the respective secondary maximum allowable transmit power  also increases since $P_{S}=P_P/2$. Thus, the secondary sum rate grows gradually until the secondary transmit power constraints in \eqref{eqn:SAP_tx_pwr} are met. At this point S-APs transmit signals with their maximum allowable transmit power $P_S$, and the secondary sum rate reaches a maximum. 
Simultaneously, the primary transmit powers at the P-APs keep increasing, and this  causes a high level of primary CCI at SUs. Consequently, the secondary sum rate decreases as $P_P$ grows without bound. When max-min transmit power allocation algorithm is adopted for both systems, the primary and secondary achievable sum rates increase with the primary transmit power $P_P$. Furthermore, Fig. \ref{fig:primary_sum_rate_vs_SNR_eql_max_min_co_loctd_DL_pilots} and Fig. \ref{fig:secondary_sum_rate_vs_SNR_eql_max_min_co_loctd_DL_pilots} reveal that the both systems achieve  higher sum rates when DL CSI  is adopted at the users for signal decoding over the statistical CSI case. In particular, the achieve rate performance of the proposed underlay spectrum sharing in cell-free massive MIMO is compared with that of the co-located counterpart in   Fig. \ref{fig:primary_sum_rate_vs_SNR_eql_max_min_co_loctd_DL_pilots} and Fig. \ref{fig:secondary_sum_rate_vs_SNR_eql_max_min_co_loctd_DL_pilots}. This comparison shows that the cell-free version outperforms the co-located case in terms of  the sum rate of the underlay spectrum sharing.  For instance, in Fig. \ref{fig:secondary_sum_rate_vs_SNR_eql_max_min_co_loctd_DL_pilots}, the cell-free based secondary system achieves   the  rate gains of $75.6\%$, $119.6\%$, and $130.0\%$  for the uniform power allocation, max-min power allocation, and DL CSI cases, respectively, compared to that of the co-located counterpart at $P_P=5$\,dBW. 

\begin{figure}[!t]\centering\vspace{-7mm}
	\includegraphics[width=0.45\textwidth]{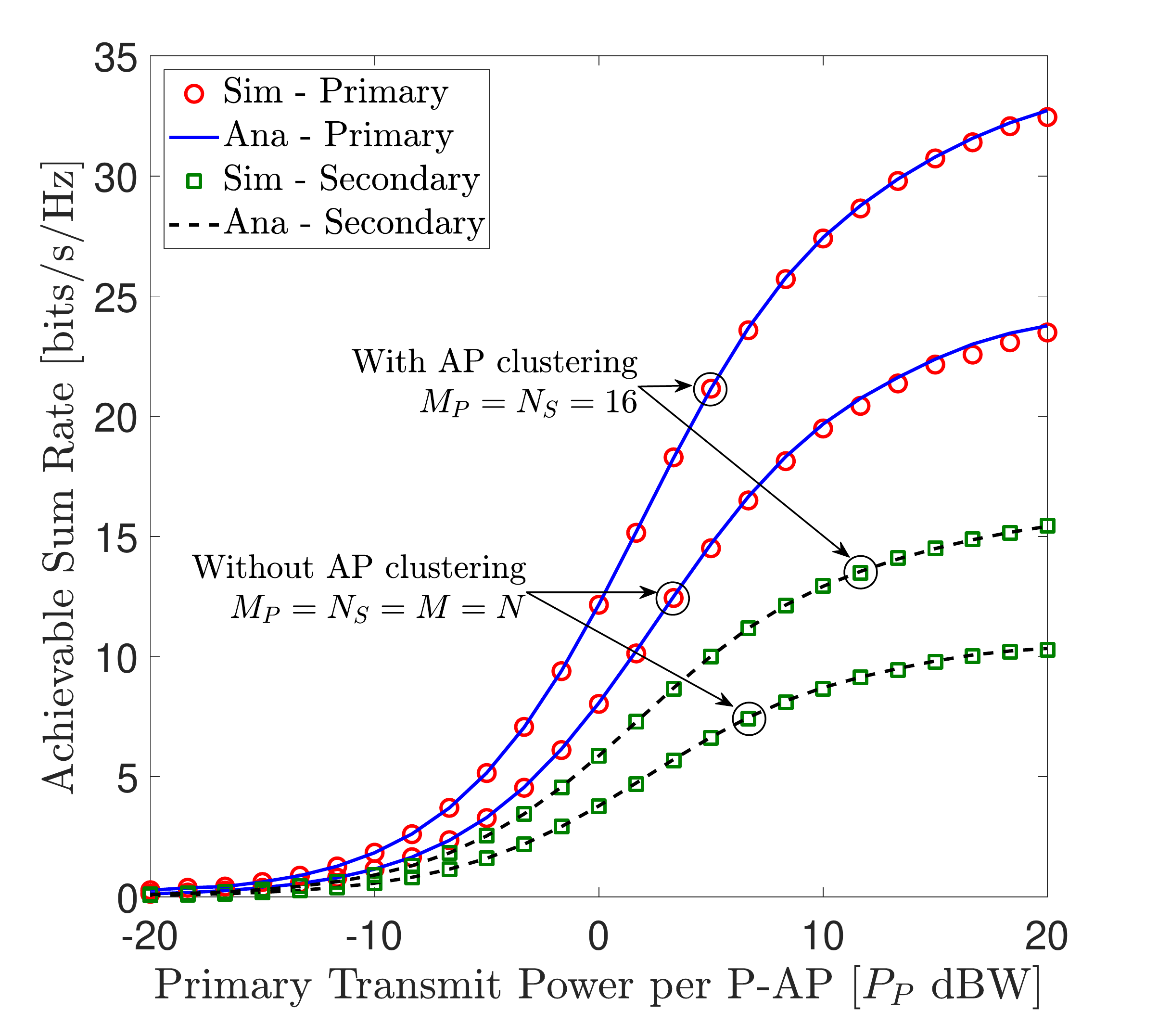}\vspace{-3mm}
	\caption{The impact of user-centric AP clustering. The achievable sum rate versus the primary transmit power with/without AP clustering for  $K = L = 10$, $M=N=32$, $I_T = 0$ dB, $\sigma_P^2 = \sigma_S^2 = 1$, $\nu = 2.4$, and $d_0 = 1$\,m. }
	\label{fig:sum_rate_USC_32_16}\vspace{-10mm}
\end{figure}

In Fig. \ref{fig:sum_rate_USC_32_16}, the effect of AP clustering is investigated. To this end the achievable sum rates of primary and secondary systems are plotted against the primary transmit power per P-AP with/without AP clustering  by adopting max-min power allocation. Fig. \ref{fig:sum_rate_USC_32_16} reveals  that AP clustering boosts the achievable rates of both   primary and secondary systems.  For example, at  $P_P$ of 10\,dBW, the primary and secondary systems achieve sum rate gains of $39.5\%$ and $48.7\%$, respectively, when  the user-centric AP clustering is adopted  over the case of uniform AP deployment without a predefined AP clustering scheme. The reason for this behavior is that, when a certain number of APs is allocated for a particular user,  the CCI from the remaining users of the own system and  the other system is  reduced. This reduced level of CCI translates into achievable rate gains. 

\begin{figure}[!t]\centering\vspace{-7mm}
	\includegraphics[width=0.45\textwidth]{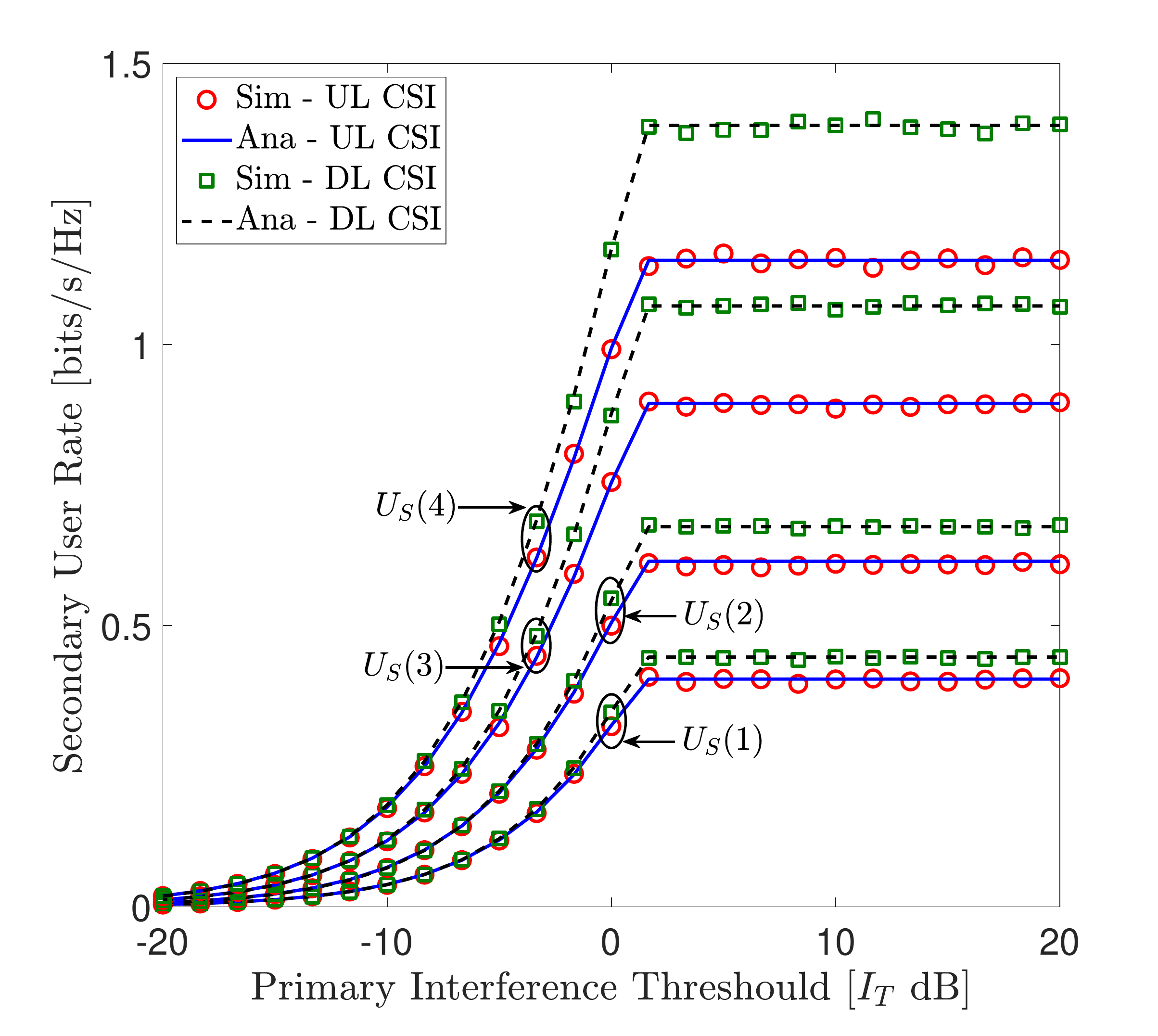}\vspace{-3mm}
	\caption{The secondary user rate versus PIT ($I_T$) for  $K = L = 4$, $M=N=32$, $P_P = 0$ dBW, $\sigma_P^2 = \sigma_S^2 = 1$, $\nu = 2.4$, and $d_0 = 1$\,m. }
	\label{fig:sum_rate_vs_Ip_32_0db_DL_pilots}\vspace{-0mm}
\end{figure}

In Fig. \ref{fig:sum_rate_vs_Ip_32_0db_DL_pilots}, the effects of secondary transmit power constraints on the  achievable secondary user rates are explored.   Two sets of rate curves are plotted as a function of PIT ($I_T$) for $M=N=32$ and with/without DL CSI at SUs  by keeping the maximum allowable transmit power $P_S=P_P/2$ at each S-AP. Fig. \ref{fig:sum_rate_vs_Ip_32_0db_DL_pilots} clearly reveals that the secondary user rates grow exponentially in the low $I_T$ regime in the both cases. In high   $I_T$ regime, the SU rates for the both CSI cases saturate to a maximum. The reason for this behavior is that in low  $I_T$ regime, a high amount of secondary CCI at  PUs is allowed, whereas in high  $I_T$ regime, the SU   rates saturate when the secondary transmit power meets the transmit power constraints in \eqref{eqn:SAP_tx_pwr}.  Fig. \ref{fig:sum_rate_vs_Ip_32_0db_DL_pilots} also reveals that the SU rates can be boosted when  the estimated DL CSI from beamformed DL pilots is used  over the statistical/long-term DL CSI counterpart for signal decoding at SUs.  For example, $U_S(3)$ and $U_S(4)$ achieve rate gains of 20.9\,\% and 18.6\,\%, respectively, by using the DL CSI at $I_T=10$\,dB compared to those of without DL CSI.

\begin{figure}[!t]\centering\vspace{-7mm}
	\includegraphics[width=0.45\textwidth]{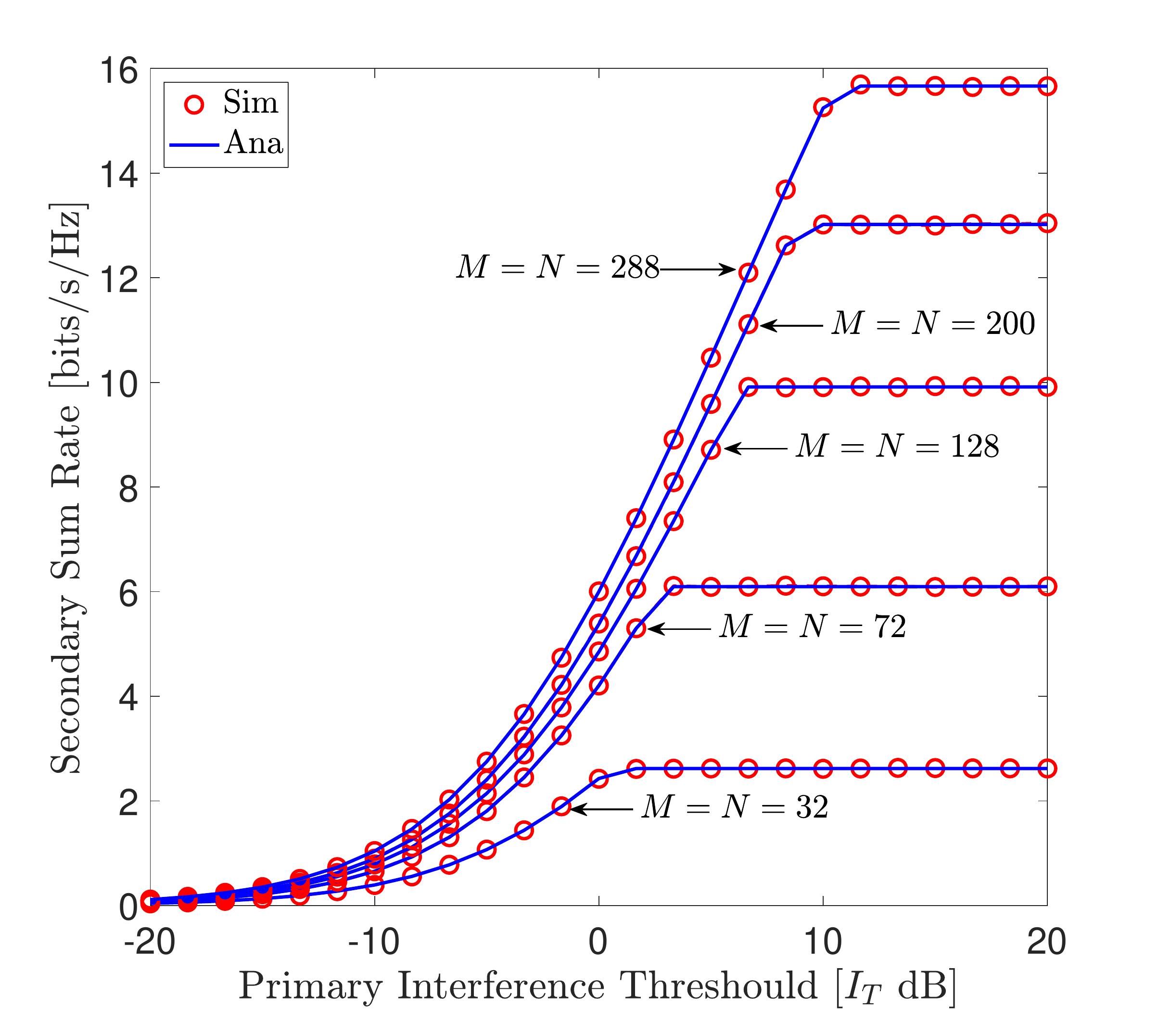}\vspace{-3mm}
	\caption{The secondary sum rate versus PIT ($I_T$) for  $K = L = 4$,  $P_P = 0$ dBW, $\sigma_P^2 = \sigma_S^2 = 1$, $\nu = 2.4$, and $d_0 = 1$\,m. }
	\label{fig:sum_rate_vs_Ip_DL_pilots_32_72_128_200_288}\vspace{-10mm}
\end{figure}

In Fig. \ref{fig:sum_rate_vs_Ip_DL_pilots_32_72_128_200_288}, the implication of the number of P-APs/S-APs is investigated by plotting  the achievable secondary sum rate  as a function of the PIT ($I_T$) with statistical DL CSI at the SUs. By varying the number of P-APs and S-APs as $M=N=32$, $M=N=72$, $M=N=128$,
$M=N=200$, and $M=N=288$, five sets of rate curves are plotted. Fig. \ref{fig:sum_rate_vs_Ip_DL_pilots_32_72_128_200_288} shows that the maximum saturation of the secondary sum rate is heavily depend on $N/M$.  The reason for this behavior can be described as follows: 
In the low $I_T$ regime, the secondary only can transmit smaller powers because  PUs can withstand only to very small level of secondary CCI. Thus, those low secondary transmit powers result in smaller secondary achievable rates.  Moreover, when $I_T$ grows large, it allows to inflect a high amount of  secondary CCI at the PUs, and thus, the S-APs can transmit high power and it exponentially increases the secondary sum rate. Once the secondary power constraints in  \eqref{eqn:SAP_tx_pwr} are met, the secondary sum rate saturates to a maximum.  

\begin{figure}[!t]\centering\vspace{-7mm}
	\includegraphics[width=0.45\textwidth]{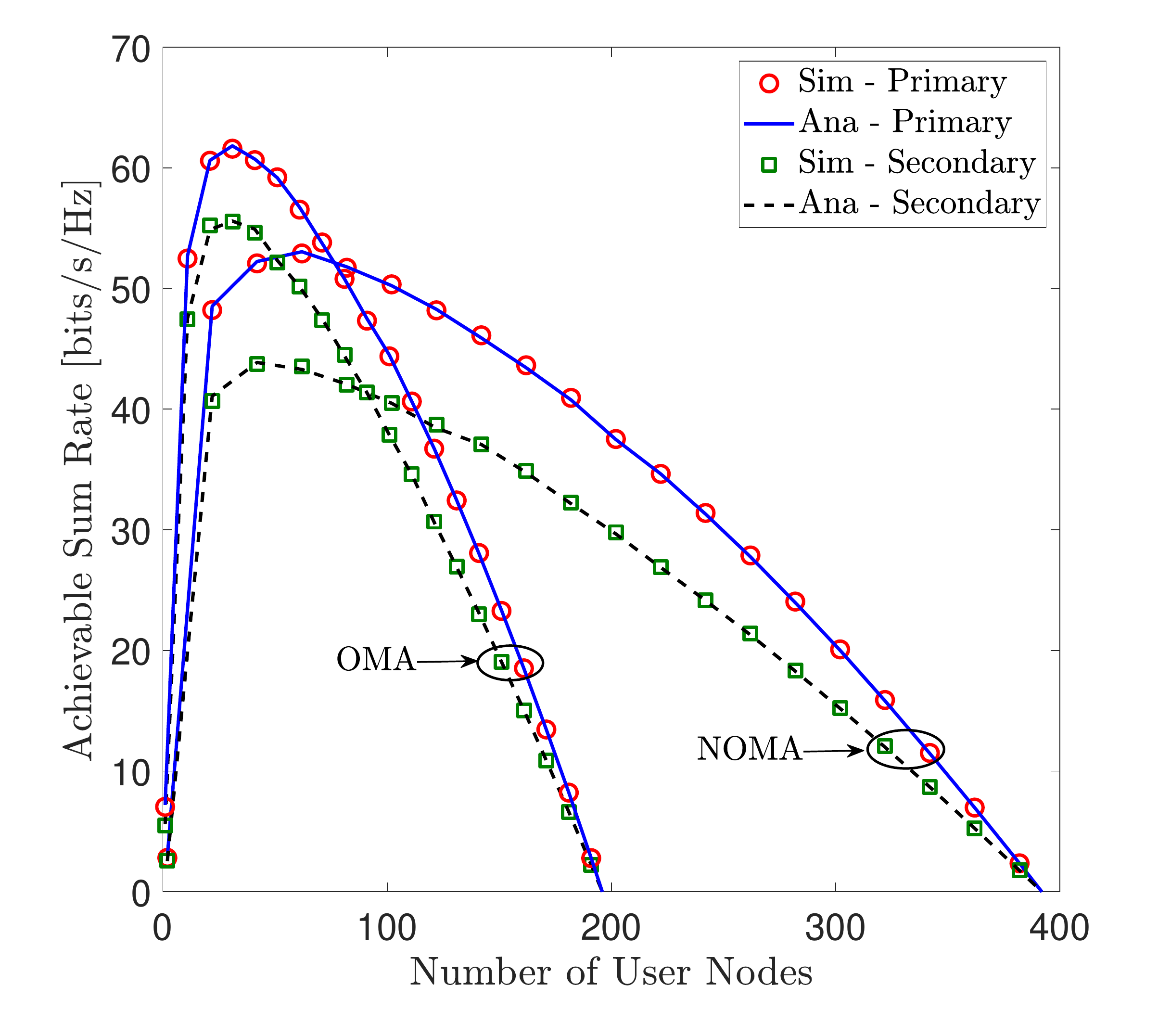}\vspace{-3mm}
	\caption{A  comparison between the achievable sum rate of NOMA and OMA for  $M = N = 32$, $P_P = 0$ dBW, $\sigma_P^2 = \sigma_S^2 = 1$, $\nu = 2.4$, $\vartheta_{P_{ai}}= \vartheta_{S_{bj}}= 0.2$, and $d_0 = 1$\,m. }
	\label{fig:rate_vs_No_US_NOMA_OMA}\vspace{-10mm}
\end{figure}

In Fig. \ref{fig:rate_vs_No_US_NOMA_OMA}, a comparison of OMA versus NOMA in terms of  the achievable sum rates of the primary and secondary systems is presented.  To this end, the sum rates are plotted as a function of the number of users that can be served simultaneously.  The analytical curves are plotted by using \eqref{eqn:noma_rate_P_sum} and \eqref{eqn:noma_rate_S_sum} for primary and secondary systems, respectively. Since the coherence interval $\tau_{c} = 196$, the maximum number of users that can be served by OMA is limited to 196. On the other hand, NOMA can serve more number of users than that of OMA  because of user clustering. However, Fig. \ref{fig:rate_vs_No_US_NOMA_OMA} reveals that OMA outperforms NOMA in the regime of low number of users. The reason for this behavior is that the intra-cluster pilot contamination due to the shared pilots among NOMA clusters and the residual interference caused by error propagation from imperfect SIC hinder the achievable rates  of NOMA in low user regime. However, NOMA-aided cell-free underlay spectrum sharing is beneficial in boosting the number of  simultaneous served SUs and the achievable sum rates at high user rate regime.  

\begin{figure}[!t]\centering\vspace{-7mm}
	\includegraphics[width=0.45\textwidth]{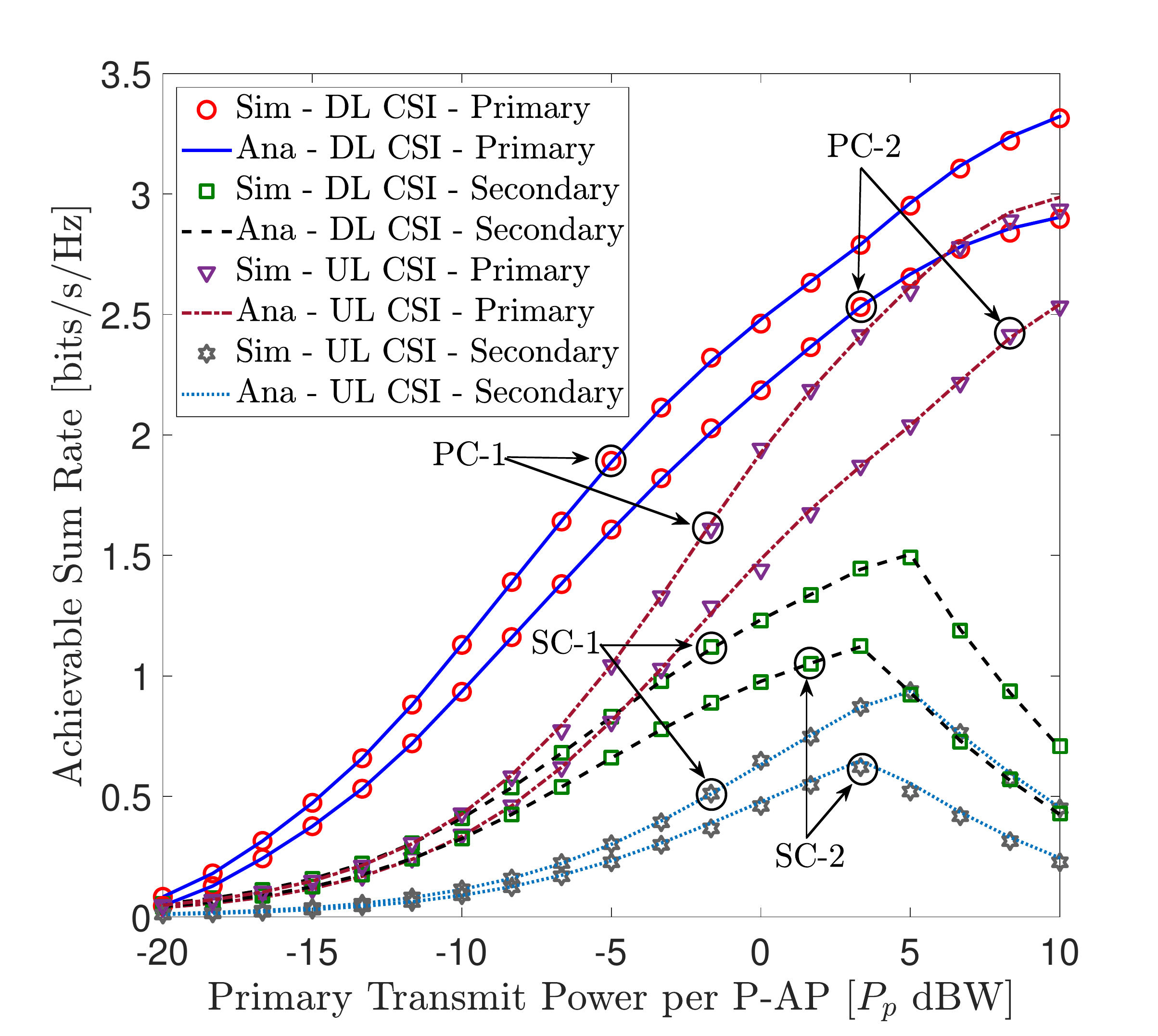}\vspace{-3mm}
	\caption{The impact of DL pilots in NOMA. The achievable sum rate versus the primary transmit power for  $M = N = 32$, $A = B = 2$, $K = L = 2$, $I_T = 0 \, dB$,   $\sigma_P^2 = \sigma_S^2 = 1$, $\nu = 2.4$,  and $d_0 = 1$\,m. Moreover, $P_S=P_P/2$ and $P_{S_n} = \min{\left(P_{S}, {I_{T_1}}\big/{Z_{1}}, \cdots, {I_{T_k}}\big/{Z_{k}}, \cdots, {I_{T_K}}\big/{Z_{K}} \right)}.$}
	\label{fig:rate_NOMA_DL_pilots}\vspace{-10mm}
\end{figure}

In Fig. \ref{fig:rate_NOMA_DL_pilots}, the impact of   DL pilots on the achievable sum rate of  NOMA-aided underlay spectrum sharing for cell-free massive MIMO is studied.
The analytical rate curves for NOMA with estimated  DL CSI  are plotted via   our analysis in  \eqref{eqn:P_SINR_DL_noma_ana} and \eqref{eqn:S_SINR_DL_noma_ana}. Fig. \ref{fig:rate_NOMA_DL_pilots} shows that the achievable sum rate of the primary system with/without DL pilots grows monotonically with $P_P$. 
Nevertheless, the sum rate of the secondary system with/without DL channel training gradually grows until a maximum in the low power regime, and then, it starts to decrease as $P_P$ continues to grow large. This is because the secondary transmit power reaches its  maximum limit and also due to higher levels of primary CCI at SUs as per the description of   Fig. \ref{fig:secondary_sum_rate_vs_SNR_eql_max_min_co_loctd_DL_pilots}.
Moreover,  Fig. \ref{fig:rate_NOMA_DL_pilots} reveals that the achievable rates can be boosted when the users adopt the estimated DL CSI via beamformed  pilots for signal decoding at PUs/SUs. For instance, at a primary transmit power of 0\,dBW, the primary system with DL pilots achieves a sum rate gain of 52.4\% over a system that only relies on statistical DL CSI (without DL pilots) for signal decoding at the PUs.

\section{Conclusion}\label{sec:conclusion}
\vspace{-1mm}
The practical feasibility of deploying underlay spectrum sharing in cell-free massive MIMO has been investigated by adopting UL/DL pilot-based channel estimations, max-min based MOOP, OMA/NOMA and the corresponding  achievable rates for both primary and secondary systems. User fairness has been guaranteed by adopting an MOOP-based max-min fairness algorithm for  P-APs/S-APs, while satisfying the secondary transmit power constraints, which are subjected to PIT. The achievable rates for both systems have been derived for locally estimated UL CSI at P-APs/S-APs and DL CSI at PUs/SUs. The impact of DL channel estimation at PUs/SUs via   beamforming of DL pilots to boost the achievable rates has been studied. We reveal that our proposed MOOP-based max-min transmit power control algorithm can significantly  boost the achievable rate of the both systems over the uniform power allocation by using carefully designed secondary transmit power constraints at each S-AP. The effect of user-centric cell-free massive MIMO in which the P-APs/S-APs are clustered to serve a particular PU/SU has been investigated. The trade-offs between OMA and NOMA in terms of the number of concurrently served PUs/SUs and the achievable rates have been explored, and thereby, it has been revealed that NOMA-aided underlay spectrum sharing can be beneficial in satisfying future massive access demands in a cell-free set-up.  
It has been shown that the achievable rates can be boosted when the  PUs/SUs estimate DL channels  from the beamformed DL pilots and adopt this estimated DL CSI to decode signals  instead of solely relying on statistical DL CSI.
Thus, our performance analysis establishes that a secondary system can be operated within the same  primary/licensed spectrum without hindering the primary system performance in a cell-free massive MIMO set-up.

\appendices
\section{}
\subsection{Derivation of the MMSE estimates in \eqref{eqn:estimate_f_mk} and \eqref{eqn:estimate_g_nl}} \label{app:Appendix1}
By substituting (\ref{eqn:est_signl_m_PAP}) into (\ref{eqn:estimate_f_mk}), $\hat{f}_{mk} $ can be derived as \cite{Kay1993}
\vspace{-1mm}
\begin{eqnarray}\label{eqn:identity_1}
	\hat{f}_{mk} &=& \frac{\E { \left( \sqrt{\tau_p P} \conj{f}_{mk} + \sqrt{\tau_p P} \conj{v}_{mk}  +  {n}_{P_m}^* \right) f_{mk} }}{\E {|\sqrt{\tau_p P} f_{mk} + \sqrt{\tau_p P} v_{mk} + {n}_{P_m}|^2 }} y_{P_{mk}} \nonumber \\
	&=& \frac{\sqrt{\tau_p P} \E{\conj{f}_{mk} {f}_{mk} } }{\tau_{p} P \E{|f_{mk}|^2} + \tau_{p} P \E{|v_{mk}|^2} + \E{|{n}_{P_m}|^2}  }  y_{P_{mk}} \stackrel{(a)}{=} c_{P_{mk}} y_{P_{mk}},
\end{eqnarray}	

\vspace{-1mm}
\noindent where $c_{P_{mk}}$ is defined in \eqref{eqn:cpmk} and ${n}_{P_m} =  \mathbf n_{P_m} \boldsymbol{\phi}_{P_k}^H$. The step $(a)$ is written by using the fact that $f_{mk}$, $v_{mk}$, and $n_{P_m}$ are Gaussian random variables with zero means and then, by evaluating  the expectations of their squared norms \cite{Cramer1970}.
Similarly, the MMSE estimate of $g_{nl}$ can be derived as given in (\ref{eqn:estimate_g_nl}).

\subsection{Derivation of $Z_{k}$ in \eqref{eqn:Z_k}} \label{app:Appendix2}
By substituting (\ref{eqn:estimate_g_nl}) and (\ref{eqn:est_signl_n_SAP}) into \eqref{eqn:rx_interf_power}, $Z_k$ is computed as
\vspace{-1mm}
\begin{eqnarray}\label{eqn:identity_1_2}
	Z_{k} &=&  \sum\nolimits_{n=1}^{N} \delta_{S_{nk}} \eta_{S_{nk}} c_{S_{nk}}^2  \E { \left|  {u_{nk}  \left( \sqrt{P_p} g_{nk} + \sqrt{P_p} u_{nk}  +     n_{S_n} \right)^* } \right|^2} \nonumber\\
	&&+  \sum\nolimits_{n=1}^{N}  \sum\nolimits_{l\neq k}^{L} \delta_{S_{nl}} \eta_{S_{nl}}  c_{S_{nl}}^2 \E { \left|  {u_{nk}  \left(  \sqrt{P_p} g_{nl} + \sqrt{P_p} u_{nl} +    n_{S_n}  \right)^*  } \right|^2} \nonumber\\
	&=&  \sum\nolimits_{n=1}^{N} \! \delta_{S_{nk}} \eta_{S_{nk}} c_{S_{nk}}^2  \zeta_{{u}_{nk}} \!\left({P_p} \!\left( \zeta_{{g}_{nk}} \!+\! 2 \zeta_{{u}_{nk}}\right)   \!+\! 1 \right) \!+\!  \sum\nolimits_{n=1}^{N} \sum\nolimits_{l \neq k}^{L} \! \delta_{S_{nl}} \eta_{S_{nl}} c_{S_{nl}}^2  \zeta_{{u}_{nk}} \! \left({P_p} \left( \zeta_{{g}_{nl}} \!+\!  \zeta_{{u}_{nl}}\right)   \!+\! 1 \right) \nonumber\\
	&=&\sum\nolimits_{n=1}^{N} \sum\nolimits_{l=1}^{L} \delta_{S_{nl}} \eta_{S_{nl}} \rho_{g_{nl}} \zeta_{{u}_{nk}} + \sum\nolimits_{n=1}^{N} \delta_{S_{nk}} \eta_{S_{nk}} \rho_{u_{nk}}^2.
\end{eqnarray}

\subsection{Derivation of the MMSE estimate of $\mu_{P_{kk}}$ in \eqref{eqn:a_pkk_hat_anly}}  \label{app:Appendix1_DL}
By noting that $f_{mk} = \hat{f}_{mk} + \epsilon_{f_{mk}}$, $\E{\mu_{P_{kk}}}$ can be calculated as
\vspace{-1mm}
\begin{eqnarray}\label{eqn:identity_a_pkk}
	\!\!\!\! \E{\mu_{P_{kk}}} &=& \E{\sum\nolimits_{m=1}^{M} \delta_{P_{mk}} \eta_{P_{mk}}^{1/2} \hat{f}_{mk} \hat{f}_{mk}^* } + \E{\sum\nolimits_{m=1}^{M} \delta_{P_{mk}} \eta_{P_{mk}}^{1/2}  \epsilon_{f_{mk}} \hat{f}_{mk}^* } = \sum\nolimits_{m=1}^{M} \delta_{P_{mk}} \eta_{P_{mk}}^{1/2} \rho_{f_{mk}},
\end{eqnarray}

\vspace{-1mm}
\noindent where $\rho_{f_{mk}} \triangleq \sqrt{\tau_{p} P} c_{P_{mk}} \zeta_{{f}_{mk}}$. Then, $\E{\lambda_{P_{kk}}}$ can be evaluated as
\vspace{-1mm}
\begin{eqnarray}\label{eqn:identity_b_pkk}
	 \E{\lambda_{P_{kk}}}  &=& \E{\sum\nolimits_{n=1}^{N} \delta_{S_{nk}} \eta_{S_{nk}}^{1/2} u_{nk}  \hat{g}_{nk}^* } \stackrel{(b)}{=} \E{ \sum\nolimits_{n=1}^{N} \delta_{S_{nk}} \eta_{S_{nk}}^{1/2} c_{S_{nk}} u_{nk} \left(\sqrt{\tau_{p} P} \left(\hat{g}_{nk}^* + \hat{u}_{nk}^*\right) + n_{S_n}  \right)  }  \nonumber \\
	&=& \sum\nolimits_{n=1}^{N} \delta_{S_{nk}} \eta_{S_{nk}}^{1/2} \rho_{u_{nk}},
\end{eqnarray}

\vspace{-1mm}
\noindent where the step $(b)$ is written by using \eqref{eqn:estimate_g_nl} and \eqref{eqn:est_signl_n_SAP}.
\noindent The variance term of $\mu_{P_{kk}}$ in \eqref{eqn:a_pkk_hat} is given by
\vspace{-1mm}
\begin{eqnarray}\label{eqn:var_a_kk}
	\Var{\mu_{P_{kk}}} = \E{|\mu_{P_{kk}}|^2} - |\E{\mu_{P_{kk}}}|^2,
\end{eqnarray}

\vspace{-1mm}
\noindent where the first expectation term in \eqref{eqn:var_a_kk} can be derived as
\vspace{-1mm}
\begin{eqnarray}\label{eqn:var_a_kk_1}
	&&\E{|\mu_{P_{kk}}|^2} =
	\E{\left|\sum\nolimits_{m=1}^{M} \delta_{P_{mk}} \eta_{P_{mk}}^{1/2} \hat{f}_{mk} \hat{f}_{mk}^* \right|^2 } + \E{\left|\sum\nolimits_{m=1}^{M} \delta_{P_{mk}} \eta_{P_{mk}}^{1/2} \epsilon_{f_{mk}} \hat{f}_{mk}^* \right|^2 } \nonumber \\
	&&=
	\!\!\sum\nolimits_{m=1}^{M} \!\!\delta_{P_{mk}} \eta_{P_{mk}} \E{\!|\hat{f}_{mk}|^4\!} \!\!+\!\! \sum\nolimits_{m=1}^{M}\! \sum\nolimits_{m' \neq m}^{M} \!\!\delta_{P_{mk}} \eta_{P_{mk}}^{1/2} \delta_{P_{m'k}} \eta_{P_{m'k}}^{1/2} \E{\!|\hat{f}_{mk}|^2 |\hat{f}_{m'k}|^2\!}  \nonumber \\
	&&\quad+
	\sum\nolimits_{m=1}^{M} \!\!\delta_{P_{mk}} \eta_{P_{mk}} \!\!\left(\zeta_{{f}_{mk}} \!-\! \rho_{f_{mk}}\right) \rho_{f_{mk}} \nonumber \\
	&&=
	\! 2 \!\! \sum\nolimits_{m=1}^{M}  \!\!\! \delta_{P_{mk}} \eta_{P_{mk}} \rho_{f_{mk}}^2 \!\!+\!\! \sum\nolimits_{m=1}^{M} \! \sum\nolimits_{m' \neq m}^{M} \!\!\! \delta_{P_{mk}} \eta_{P_{mk}}^{1/2} \delta_{P_{m'k}} \eta_{P_{m'k}}^{1/2} \rho_{f_{mk}} \rho_{f_{m'k}} \!\!+\!\! \sum\nolimits_{m=1}^{M}  \!\!\! \delta_{P_{mk}} \eta_{P_{mk}} \!\left(\zeta_{{f}_{mk}} \!\!-\! \rho_{f_{mk}}\right) \!\rho_{f_{mk}} \nonumber \\
	&&=
	\sum\nolimits_{m=1}^{M} \sum\nolimits_{m'=1}^{M} \delta_{P_{mk}} \eta_{P_{mk}}^{1/2} \delta_{P_{m'k}} \eta_{P_{m'k}}^{1/2} \rho_{f_{mk}} \rho_{f_{m'k}} + \sum\nolimits_{m=1}^{M} \delta_{P_{mk}} \eta_{P_{mk}} \zeta_{{f}_{mk}} \rho_{f_{mk}}.
\end{eqnarray}

\vspace{-1mm}
\noindent By substituting \eqref{eqn:identity_a_pkk} and \eqref{eqn:var_a_kk_1} into \eqref{eqn:var_a_kk}, we have
\vspace{-1mm}
\begin{eqnarray}\label{eqn:var_a_kk_3}
	\Var{\mu_{P_{kk}}} = \sum\nolimits_{m=1}^{M} \delta_{P_{mk}} \eta_{P_{mk}} \zeta_{{f}_{mk}} \rho_{f_{mk}} \triangleq v_{P_{kk}}.
\end{eqnarray}

\vspace{-1mm}
\noindent The variance of $\lambda_{P_{kk}}$ can be given as
\vspace{-1mm}
\begin{eqnarray}\label{eqn:var_b_pkk}
	\Var{\lambda_{P_{kk}}} = \E{|\lambda_{P_{kk}}|^2} - |\E{\lambda_{P_{kk}}}|^2.
\end{eqnarray}

\vspace{-1mm}
\noindent Thus, the first expectation term in \eqref{eqn:var_b_pkk} can be derived as
\vspace{-1mm}
\begin{eqnarray}\label{eqn:var_b_kk_1}
	\E{|\lambda_{P_{kk}}|^2} &=& \E{\left|\sum\nolimits_{n=1}^{N} \delta_{S_{nk}} \eta_{S_{nk}}^{1/2} {u}_{nk} \hat{g}_{nk}^* \right|^2 } 
	\stackrel{(c)}{=} \sum\nolimits_{n=1}^{N} \delta_{S_{nk}} \eta_{S_{nk}} c_{S_{nk}}^{2} \zeta_{{u}_{nk}} \left( \tau_{p} P \left( \zeta_{{g}_{nk}} +2 \zeta_{{u}_{nk}} \right) +1\right) \nonumber \\
	&=& \sum\nolimits_{n=1}^{N} \delta_{S_{nk}} \eta_{S_{nk}} \rho_{g_{nk}} \zeta_{{u}_{nk}} + \sum\nolimits_{n=1}^{N} \delta_{S_{nk}} \eta_{S_{nk}} \rho_{u_{nk}}^2, 
\end{eqnarray}

\vspace{-1mm}
\noindent where the step $(c)$ is written by following steps similar to those used in the step $(b)$ in \eqref{eqn:identity_b_pkk}. Therefore, by using the results in \eqref{eqn:identity_b_pkk} and \eqref{eqn:var_b_kk_1}, we have
\vspace{-1mm}
\begin{eqnarray}\label{eqn:var_b_pkk_3}
	\Var{\lambda_{P_{kk}}} = \sum\nolimits_{n=1}^{N} \delta_{S_{nk}} \eta_{S_{nk}} \zeta_{{u}_{nk}} \rho_{g_{nk}} \triangleq u_{P_{kk}}.
\end{eqnarray} 

\vspace{-1mm}
\noindent Then, by using \eqref{eqn:identity_a_pkk}, \eqref{eqn:identity_b_pkk} , \eqref{eqn:var_a_kk_3}, and \eqref{eqn:var_b_pkk_3}, the effective channel estimate can be derived as \eqref{eqn:a_pkk_hat_anly} \cite{Interdonato2016}.

\section{Derivation of SINR in (\ref{eqn:SINR_ k_PU}) } \label{app:Appendix3}
The expectation   in   numerator of (\ref{eqn:SINR_ k_PU}) can be derived   as  
\vspace{-1mm}
\begin{eqnarray}\label{eqn:identity_2}	
	\E {\sum\nolimits_{m=1}^{M} \delta_{P_{mk}} {\eta_{P_{mk}}^{1/2} f_{mk} \hat{f}_{mk}^*}}  = \sum\nolimits_{m=1}^{M} \eta_{P_{mk}}^{1/2} \E{\left(\hat{f}_{mk} + \epsilon_{f_{mk}} \right) \conj{f}_{mk} } \stackrel{(d)}{=}  \sum\nolimits_{m=1}^{M} \delta_{P_{mk}} \eta_{P_{mk}}^{1/2} \rho_{f_{mk}},
\end{eqnarray}

\vspace{-1mm}
\noindent where $\epsilon_{f_{mk}}$ is an estimation error of $f_{mk}$ such that  $f_{mk} = \hat{f}_{mk}+\epsilon_{f_{mk}} $, satisfying $\E{\epsilon_{f_{mk}} \conj{f}_{mk}}=0 $. In \eqref{eqn:identity_2}, the step $(d)$ is written by substituting (\ref{eqn:estimate_f_mk}) and then evaluating   expectation term as 
\vspace{-1mm}
\begin{eqnarray}\label{eqn:identity_3}
	\E{|\hat{f}_{mk}|^2} = c_{P_{mk}}^2 \E{| y_{P_{mk}}|^2}=\sqrt{\tau_{p} P}\zeta_{f_{mk}} c_{P_{mk}}= \rho_{f_{mk}}.
\end{eqnarray}

\vspace{-1mm}
\noindent Then, the variance term  in (\ref{eqn:SINR_ k_PU}) can be derived as
\vspace{-1mm}
\begin{eqnarray}\label{eqn:identity_4}
	\Var{\sum\nolimits_{m=1}^{M} \delta_{P_{mk}} {\eta_{P_{mk}}^{1/2} f_{mk} \hat{f}_{mk}^*}}
	&& 
	= \sum\nolimits_{m=1}^{M}  \delta_{P_{mk}} \eta_{P_{mk}} \left( \E{\left|\left( \hat{f}_{mk} +  \epsilon_{f_{mk}}  \right)  \conj{\hat{f}}_{mk}\right|^2}   - \rho_{f_{mk}}^2 \right) \nonumber\\
	&& \!\!\!\!\!\!\!\!\!\!\!\!\!\!\!\!\!\!\!\!\!\!\!\!\!\!\!\!\!\!\!\!\!\!\!\!\!\!\!\!\!\!\!\!\!\!\!\!\!\!\!\!\!\!\!\!\!\!\!\!\!\!
	=   \sum\nolimits_{m=1}^{M}  \delta_{P_{mk}} \eta_{P_{mk}}  \left( 2\rho_{f_{mk}}^2  +  \rho_{f_{mk}}\left(\zeta_{f_{mk}}  -  \rho_{f_{mk}} \right)  -  \rho_{f_{mk}}^2 \right)   
	=  \sum\nolimits_{m=1}^{M} \delta_{P_{mk}} \eta_{P_{mk}} \rho_{f_{mk}} \zeta_{f_{mk}}. 
\end{eqnarray}

\vspace{-1mm}
\noindent The expectation of the first term in \eqref{eqn:Ip2} can be computed as
\vspace{-1mm}
\begin{eqnarray}\label{eqn:identity_5}
	\!\!\!\!\!\!\!\E {\left|\sum\nolimits_{i \neq k}^{K}{\sum\nolimits_{m=1}^{M} \delta_{P_{mi}} {\eta_{P_{mi}}^{1/2} f_{mk} \hat{f}_{mi}^* } }\right|^2 }
	&& 
	\stackrel{(e)}{=}  \sum\nolimits_{i \neq k}^{K} \sum\nolimits_{m=1}^{M} \E{\left| \delta_{P_{mi}} \eta_{P_{mi}}^{1/2} f_{mk} c_{P_{mi}} y_{P_{mi}} \right|^2 } \nonumber\\
	&& \!\!\!\!\!\!\!\!\!\!\!\!\!\!\!\!\!\!\!\!\!\!\!\!\!\!\!\!\!\!\!\!\!\!\!\!\!\!\!\!\!\!\!\!\!\!\!\!\!\!\!\!\!\!\!\!\!\!\!\!\!\!\!\!\!\!\!\!\!\!\!\!\!\!\!\!\!\!\!\!\!\!\!\!\!\!\!\!\!\!\!
	\stackrel{(f)}{=}   \sum\nolimits_{i \neq k}^{K}  \sum\nolimits_{m=1}^{M}  \delta_{P_{mi}} \eta_{P_{mi}} c_{P_{mi}}^{2} \zeta_{f_{mk}} \left( \tau_{p} P  \left( \zeta_{f_{mi}}  + \zeta_{v_{mi}}\right) + 1 \right) 
	=\sum\nolimits_{i\neq k}^{K} \sum\nolimits_{m=1}^{M} \delta_{P_{mi}} \eta_{P_{mi}} \rho_{f_{mi}} \zeta_{{f}_{mk}},
\end{eqnarray}

\vspace{-1mm}
\noindent where the steps $(e)$ and   $(f)$ are written by using \eqref{eqn:estimate_f_mk} and \eqref{eqn:est_signl_m_PAP}, respectively. 
The expectation in the second term in \eqref{eqn:Ip2} can be derived by following steps similar to those used in  \eqref{eqn:identity_1_2} as follows:
\vspace{-1mm}
\begin{eqnarray}\label{eqn:identity_6}
	\!\!\!\!\!\!\!\!\!\!\!\!	\E {\left|\sum\nolimits_{j=1}^{L} {\sum\nolimits_{n=1}^{N} \delta_{S_{nj}} {\eta_{S_{nj}}^{1/2} u_{nk} \hat{g}_{nj}^* } }\right|^2 }
	 &=& 
	\sum_{n=1}^{N} \delta_{S_{nk}} \eta_{S_{nk}} c_{S_{nk}}^2  \zeta_{u_{nk}} \left( \tau_{p} P\left(\zeta_{g_{nk}}   +   2 \zeta_{u_{nk}}\right)+1 \right) \nonumber\\
	&&
	+  \sum\nolimits_{j \neq k}^{L}   \sum\nolimits_{n=1}^{N} \delta_{S_{nj}}  \eta_{S_{nj}} c_{S_{nj}}^2  \zeta_{u_{nk}} \left( \tau_{p} P\left( \zeta_{g_{nj}}   +    \zeta_{u_{nj}} \right)+1\right)  \nonumber \\
	&=&
	 \sum\nolimits_{j=1}^{L} \sum\nolimits_{n=1}^{N} \delta_{S_{nj}} \eta_{S_{nj}} \rho_{g_{nj}} \zeta_{{u}_{nk}} + \sum\nolimits_{n=1}^{N} \delta_{S_{nk}} \eta_{S_{nk}} \rho_{u_{nk}}^2.
\end{eqnarray}

\vspace{-1mm}
\noindent By substituting (\ref{eqn:identity_2}), (\ref{eqn:identity_4}),  (\ref{eqn:identity_5}), and (\ref{eqn:identity_6}) into (\ref{eqn:SINR_ k_PU}), the desired SINR   can be derived as in (\ref{eqn:sinr_k_PU}).

\section{Derivation of SINR in (\ref{eqn:P_SINR_DL_rearngd_E_anly}) } \label{app:Appendix4}
The expectation of   magnitude squared error $(\epsilon^{\mu}_{P_{kk}})$ in   \eqref{eqn:P_SINR_DL_rearngd_E} can be calculated as
\vspace{-1mm}
\begin{eqnarray}\label{eqn:identity_7}
	\E{|\epsilon^{\mu}_{P_{kk}}|^2} &=& \E{|\mu_{P_{kk}} - \hat{\mu}_{P_{kk}} |^2} 
    \stackrel{(g)}{=}
	\frac{\left(1+P_{p,d} u_{P_{kk}}\right)^2 \Var{\mu_{P_{kk}}} + \left(P_{p,d} v_{P_{kk}}\right)^2 \Var{\lambda_{P_{kk}}} + P_{p,d} v_{P_{kk}}^2 }{\left(P_{p,d} \left(v_{P_{kk}} + u_{P_{kk}}\right) +1\right)^2} \nonumber\\
	&=&
	\frac{\left(1+P_{p,d} u_{P_{kk}}\right)^2 v_{P_{kk}} + \left(P_{p,d} v_{P_{kk}}\right)^2 u_{P_{kk}} + P_{p,d} v_{P_{kk}}^2 }{\left(P_{p,d} \left(v_{P_{kk}} + u_{P_{kk}}\right) +1\right)^2} 
	\kappa^{\epsilon}_{P_{kk}},
\end{eqnarray} 

\vspace{-1mm}
\noindent where $\kappa^{\epsilon}_{P_{kk}}$ is defined in the first term of \eqref{eqn:kappa_v_Pki} and the step $(g)$ is written by substituting \eqref{eqn:a_pkk_hat_anly} and then, evaluating the expectation.
The expectation term of $\mu_{P_{ki}}$ in the denominator of \eqref{eqn:P_SINR_DL_rearngd_E} can be derived as
\vspace{-1mm}
\begin{eqnarray}\label{eqn:E_a_ki}
	\!\!\!	\!\!\!\!\!\!\!\!\!\! \E{|\mu_{P_{ki}} |^2} \!&=&\! \E{\left|\sum\nolimits_{m=1}^{M} \!\delta_{P_{mi}} \eta_{P_{mi}}^{1/2} \hat{f}_{mk} \hat{f}_{mi}^* \right|^2 } \!\! + \! \E{\left|\sum\nolimits_{m=1}^{M} \! \delta_{P_{mi}} \eta_{P_{mi}}^{1/2} \epsilon_{f_{mk}} \hat{f}_{mi}^* \right|^2 }  \nonumber \\
	&=&
	\sum\nolimits_{m=1}^{M} \! \delta_{P_{mi}} \eta_{P_{mi}} \zeta_{f_{mk}} \rho_{f_{mi}} \!\triangleq\! v_{P_{ki}}.
\end{eqnarray} 

\vspace{-1mm}
\noindent Then, the expectation term  in  \eqref{eqn:P_SINR_DL_rearngd_E} with $\lambda_{P_{kj}}$ is evaluated similar to the steps those used in \eqref{eqn:identity_6} as
\vspace{-1mm}
\begin{eqnarray}\label{eqn:E_b_pkj}
	\!\!\!\!\!\!\sum\nolimits_{j=1}^{L}\E{|\lambda_{P_{kj}} |^2} \!= \! \sum\nolimits_{j=1}^{L} \sum\nolimits_{n=1}^{N} \delta_{S_{nj}} \eta_{S_{nj}} \rho_{g_{nj}} \zeta_{{u}_{nk}} \! + \! \sum\nolimits_{n=1}^{N} \delta_{S_{nk}} \eta_{S_{nk}} \rho_{u_{nk}}^2.
\end{eqnarray} 

\vspace{-1mm}
\noindent The expectation term in the numerator of   \eqref{eqn:P_SINR_DL_rearngd_E} is given by
\vspace{-1mm}
\begin{eqnarray}\label{eqn:E_a_kk_hat}
	\E{|\hat{\mu}_{P_{kk}}|^2} &=& \E{|\mu_{P_{kk}} - \epsilon^{\mu}_{P_{kk}} |^2} = \E{|\mu_{P_{kk}}|^2} - \E{|\epsilon^{\mu}_{P_{kk}} |^2}.
\end{eqnarray}

\vspace{-1mm}
\noindent By substituting \eqref{eqn:var_a_kk_1} and \eqref{eqn:identity_7} into \eqref{eqn:E_a_kk_hat}, we have
\vspace{-1mm}
\begin{eqnarray}\label{eqn:E_a_kk_hat_1}
	\E{|\hat{\mu}_{P_{kk}}|^2}  =  \sum\nolimits_{m=1}^{M} \sum\nolimits_{m'=1}^{M} \delta_{P_{mk}} \eta_{P_{mk}}^{1/2} \delta_{P_{m'k}} \eta_{P_{m'k}}^{1/2} \rho_{f_{mk}} \rho_{f_{m'k}} + v_{P_{kk}}  - \kappa^{\epsilon}_{P_{kk}}.
\end{eqnarray}

\vspace{-1mm}
\noindent By substituting \eqref{eqn:identity_7}, \eqref{eqn:E_a_ki}, \eqref{eqn:E_b_pkj}, and \eqref{eqn:E_a_kk_hat_1} into \eqref{eqn:P_SINR_DL_rearngd_E}, the  effective DL SINR at $U_P(k)$ is evaluated as   \eqref{eqn:P_SINR_DL_rearngd_E_anly}. By following step similar to \eqref{eqn:identity_7}-\eqref{eqn:E_a_kk_hat_1}, the DL   SINR at $U_S(l)$ can be computed as \eqref{eqn:S_SINR_DL_rearngd_E_anly}.

\linespread{1.5}
\bibliographystyle{IEEEtran}
\bibliography{IEEEabrv,References_1}

\end{document}